\documentclass[aps,prd,superscriptaddress,floatfix,nofootinbib,showkeys,showpacs,twocolumn,reprint]{revtex4-1}
\usepackage{amsmath}
\usepackage{amssymb}
\usepackage{graphicx}
\usepackage[normalem]{ulem}
\usepackage{color}
\usepackage{xspace}
\usepackage[colorinlistoftodos]{todonotes}

\newcommand{\gray}{$\gamma$-ray\xspace}
\newcommand{\unit}[1]{\,\mathrm{#1}\xspace}

\newcommand{\FermiLAT}{\emph{Fermi}~LAT\xspace}
\newcommand{\fermiLAT}{\emph{Fermi}-LAT\xspace}
\newcommand{\vev}{m_f}
\newcommand{\Mmed}{M_{\rm med}}

\begin{document}

\title{\boldmath Sensitivity of the Cherenkov Telescope Array to the detection of a dark matter signal in comparison to direct detection and collider experiments}

\author{Csaba Bal\'azs}
\email{csaba.balazs@monash.edu}
\affiliation{ARC Centre of Excellence for Particle Physics at the Tera-scale, School of Physics and Astronomy, Monash University, Melbourne, Victoria 3800, Australia}
\affiliation{Monash Centre for Astrophysics, School of Physics and Astronomy,
Monash University, Melbourne, Victoria 3800 Australia}
\author{Jan Conrad}
\email{conrad@fysik.su.se}
\affiliation{Department of Physics, Stockholm University, AlbaNova, SE-106 91 Stockholm, Sweden}
\affiliation{The Oskar Klein Centre for Cosmoparticle Physics, AlbaNova, SE-106 91 Stockholm, Sweden}
\author{Ben Farmer}
\email{benjamin.farmer@fysik.su.se}
\affiliation{Department of Physics, Stockholm University, AlbaNova, SE-106 91 Stockholm, Sweden}
\affiliation{The Oskar Klein Centre for Cosmoparticle Physics, AlbaNova, SE-106 91 Stockholm, Sweden}
\author{Thomas Jacques}
\email{thomas.jacques@sissa.it}
\affiliation{SISSA and INFN, via Bonomea 265, 34136 Trieste, Italy}
\author{Tong Li}
\email{tong.li@monash.edu}
\affiliation{ARC Centre of Excellence for Particle Physics at the Tera-scale, School of Physics and Astronomy, Monash University, Melbourne, Victoria 3800, Australia}
\author{Manuel Meyer}
\thanks{Corresponding author}
\email{mameyer@stanford.edu}
\affiliation{W. W. Hansen Experimental Physics Laboratory,
Kavli  Institute  for  Particle  Astrophysics  and  Cosmology,
Department  of  Physics  and  SLAC  National  Accelerator  Laboratory,
Stanford  University,  Stanford,  California  94305,  USA}
\affiliation{Department of Physics, Stockholm University, AlbaNova, SE-106 91 Stockholm, Sweden}
\affiliation{The Oskar Klein Centre for Cosmoparticle Physics, AlbaNova, SE-106 91 Stockholm, Sweden}
\author{Farinaldo S. Queiroz}
\email{queiroz@mpi-hd.mpg.de}
\affiliation{Max-Planck-Institut f\"ur Kernphysik, Saupfercheckweg 1, 69117 Heidelberg, Germany}
\author{Miguel A.~S\'anchez-Conde}
\email{miguel.sanchezconde@uam.es}
\affiliation{Department of Physics, Stockholm University, AlbaNova, SE-106 91 Stockholm, Sweden}
\affiliation{The Oskar Klein Centre for Cosmoparticle Physics, AlbaNova, SE-106 91 Stockholm, Sweden}
\affiliation{Instituto de F\'{\i}sica Te\'orica UAM/CSIC, Universidad Aut\'onoma de Madrid, E-28049 Madrid, Spain}
\affiliation{Departamento de F\'isica Te\'orica, M-15, Universidad Aut\'onoma de Madrid, E-28049 Madrid, Spain}

\date{\today}

\begin{abstract}
Imaging atmospheric Cherenkov telescopes (IACTs) that are sensitive to potential $\gamma$-ray signals  from dark matter (DM) annihilation above $\sim50\,$GeV will soon be superseded by the Cherenkov Telescope Array (CTA). CTA will have a point source sensitivity an order of magnitude better than currently operating IACTs and will cover a broad energy range between 20\,GeV and 300\,TeV.  Using effective field theory and simplified models to calculate $\gamma$-ray spectra resulting from DM annihilation, we compare the prospects to constrain such models with CTA observations of the Galactic center with current and near-future measurements at the Large Hadron Collider (LHC) and direct detection experiments. For DM annihilations via vector or pseudoscalar couplings, CTA observations will be able to probe DM models out of reach of the LHC, and, if DM is coupled to standard fermions by a pseudoscalar particle, beyond the limits of current direct detection experiments. 
\end{abstract}


\maketitle

\section{Introduction}
\label{sec:intro}
Astrophysical evidence suggests that 84\,\% 
of the matter in the Universe is composed 
of cold dark matter (DM)~\cite{Ade:2015xua}.
New particles beyond the Standard Model (SM)
might constitute the entirety of DM, but the characteristics
  of such particles and their interactions with the SM remain 
unknown. 
One widely studied candidate is a \emph{weakly interacting massive 
particle} (WIMP). According to the so-called WIMP miracle, 
if DM consists of such particles 
with masses of the order of TeV and weak scale interactions, 
they could provide 
the right DM relic abundance~\cite{Srednicki:1985sf}.

A large number of experiments are searching for DM using essentially three different approaches.
Direct detection (DD) looks for recoils caused by nucleon-WIMP scattering. 
Different collaborations have used 
different target materials such as liquid xenon (XENON, LUX, or PandaX experiments~\cite{Aprile:2016swn,Akerib:2016vxi,Fu:2016ega}) or solid state detectors (Ge: CDMS, CoGeNT, NaI: DAMA~\cite{Agnese:2015nto,Aalseth:2012if,Bernabei:2016bkl}). 
See Ref.~\cite{Undagoitia:2015gya} for a recent review.
In collider searches, DM could be produced in the collisions of SM particles and manifest itself as missing energy in the final state. The ATLAS and CMS experiments at the Large Hadron Collider (LHC) continue to search for such signatures~\cite{Khachatryan:2016dvc,Tolley:2016lbg,Aaboud:2016obm,ATLAS:2016tsc,Sirunyan:2017hci,Sirunyan:2017onm}.
The third approach is indirect detection (ID)
where one searches for SM particles as a result of DM decay or
annihilation from astrophysical objects which should harbor a large 
amount of DM (we focus on DM annihilation in this work). 
Examples are the IceCube telescope, which looks for neutrinos~\cite{Aartsen:2016zhm,Aartsen:2016fep}, AMS, which measures charged cosmic rays~\cite{Bergstrom:2013jra,Giesen:2015ufa}, as well as the \emph{Fermi} Large Area Telescope (LAT) and imaging air Cherenkov telescopes (IACTs) such as 
H.E.S.S., VERITAS, and MAGIC that are sensitive to high and very high energy $\gamma$ rays, respectively~\cite{Abramowski:2011hc,Nieto:2015hca,Zitzer:2015eqa,Baring:2015sza,Ackermann:2015zua,Abdallah:2016ygi,Fermi-LAT:2016uux,Profumo:2016idl,Ahnen:2016qkx}.

To compare constraints from these different experiments and approaches, 
one has to invoke an underlying theory of the DM interaction. 
Effective field theories (EFTs) and simplified models 
provide such a framework in a generic 
way.
In EFTs, the only additional degree of freedom is the DM particle.
Any fields mediating between the DM and SM are assumed to be heavy, compared to the energy of the relevant interactions, and integrated out. 
In this way, effective operators describe
the interaction between DM and SM particles. 
The EFT approach is valid as long as the center-of-mass energy of the relevant interaction is small in 
comparison to the mass of the mediator so that the
mediator cannot be produced on shell.
This is typically a problem for collider searches and not as severe for 
ID as the velocity of DM particles in astrophysical systems is small \cite{Busoni:2013lha,Busoni:2014sya,Busoni:2014haa}. 
Where the EFT fails one can use simplified models in which 
at least one additional particle is introduced that mediates between 
the DM and SM sectors, furnishing a closer connection to UV complete models. For some recent reviews, see e.g. Refs.~\cite{Abdallah:2015ter,DeSimone:2016fbz}.

The goal of the present study is to compare the DM detection sensitivity of the
Cherenkov Telescope Array (CTA) to that of DD and collider experiments. 
With its large foreseen energy range between 20\,GeV and 300\,TeV
and a point source sensitivity a factor of 10 better than current IACTs~\cite{Consortium:2010bc}, 
ID DM searches with CTA should yield unprecedented complementary results 
to that of DD experiments and colliders~\cite{Doro:2012xx}. 
One of the most promising targets for DM searches with CTA is the Galactic center (GC) due to its relative proximity and high DM density~\cite{Doro:2012xx,Gondolo:1999ef,Cesarini:2003nr,Ullio:2001fb,Aharonian:2006wh,Hooper:2012sr}. 

The paper is outlined as follows.
In Sec.~\ref{sec:dmsig} we discuss the DM density profiles used in this study, and derive the expected DM signal 
from EFTs and simplified models, focusing on models that facilitate comparison to LHC results.
Then, we briefly discuss expected backgrounds from astrophysical sources in Sec.~\ref{sec:bkg}. We describe our analysis
framework and observational strategy in Sec.~\ref{sec:analysis}, which will yield a realistic estimate of the CTA sensitivity 
to the detection of DM. 
Finally, we present our results and comparison to DD and collider 
experiments in Sec.~\ref{sec:results}, where we also discuss the 
validity range for EFTs and simplified models.
Our conclusions are drawn in Sec.~\ref{sec:concl}.

\section{Expected dark matter signal}
\label{sec:dmsig}
The expected $\gamma$-ray flux $d\phi/dE$
from DM annihilation is given by (e.g.~\cite[][]{1998APh.....9..137B,2004PhRvD..69l3501E})
\label{sec:intro}
\begin{equation}
\frac{d\phi}{dE} = 
\frac{x}{4\pi}\frac{\langle\sigma v\rangle}{2m_\chi^2}
\sum\limits_f B_f \frac{dN_f}{dE}
\int\limits_{\Delta\Omega}\,\int\limits_\mathrm{LOS}
\rho_\chi^2(r)\,dl\,d\Omega,
\label{eq:annflux}
\end{equation}
where $\langle \sigma v \rangle$ is the 
velocity-averaged annihilation cross section, 
$m_\chi$ the DM mass, $dN_f / dE$
describes the $\gamma$-ray spectra per annihilation for the annihilation channel into SM particle $f$ with branching ratio $B_f$, 
and $x = 1$ for Majorana and $x = 1/2$ for Dirac DM, respectively. 
These spectra are calculated in the frameworks of EFTs and 
simplified models and are described below in Secs. \ref{sec:EFTtheory} and \ref{sec:simpmods}, respectively.
The double integral over the solid angle $\Delta\Omega$ and line of sight (LOS) over the squared DM energy density $\rho_\chi$ is commonly 
denoted as the astrophysical {\it J~factor}. We describe in detail our choices in computing this parameter in the next subsection.

\subsection{The astrophysical J~factor}  \label{sec:profiles}

The key ingredient in the calculation of the J~factor is the local DM density profile $\rho_\chi(r)$, which describes the way the DM is distributed in the Galaxy. Unfortunately, this is currently  poorly constrained by observations, with very large uncertainties 
 particularly in the innermost regions (e.g. Refs.~\cite{2011JCAP...11..029I,Iocco:2015xga,Pato:2015dua,2017PDU....15...90I} and references therein). Indeed, at present it is not possible to even motivate or build a model for the Milky Way (MW) DM density profile that would entirely be based on observational data alone. Instead, results from $N$-body cosmological simulations 
have been traditionally used both to propose parametric expressions of the profile and to guide our particular parameter choices for the MW. Two of the most commonly used DM density profiles are the so-called Navarro-Frenk-White (NFW) \cite{Navarro:1995iw,1997ApJ...490..493N}:
\begin{equation}
\rho_\text{NFW}(r) = \frac{\rho_0}{\left(\frac{r}{r_s}\right) \left[1 +\left(\frac{r}{r_s}\right)\right]^2}\ , 
\label{eq:NFW}
\end{equation}
where $\rho_0$ and $r_s$ represent a characteristic density and a scale radius, respectively; and the Einasto profile \cite{Einasto:1965,Navarro:2003ew}:
\begin{equation}
\rho_{\text{Ein}}(r)=\rho_0 \exp \left\{-\frac{2}{\alpha}\left[\left(\frac{r}{r_s}\right)^{\alpha} - 1\right] \right\}\ .
\label{eq:einasto}
\end{equation}

Both profiles have been shown to provide very good fits to $N$-body simulation data at all halo mass scales and cosmological epochs (e.g.~\cite[][]{Navarro:2003ew,Merritt:2005xc,Maccio':2008xb,Klypin:2010qw}). Yet, we note that these results were based on DM-only simulations and thus disregard any possible effects due to baryons. Baryonic processes such as gas dissipation, star formation, and supernova feedback are expected to be particularly relevant at the centers of galaxies like our own, where baryons represent indeed the dominant gravitational component \cite{Iocco:2015xga}. The precise impact of this ordinary matter on the DM density profile remains unclear at present (e.g.~\cite[][]{Colin:2005rr,Gustafsson:2006gr,Mashchenko:2007jp,Tissera:2009cm,Governato:2009bg,SommerLarsen:2009me,Gnedin:2011uj,
Pontzen:2011ty,Maccio':2011eh,2014MNRAS.437..415D,2016MNRAS.455.4442S}).

  As will be explained below, in this work, we will focus on regions around the GC and, thus, the inner DM density profile of the MW becomes particularly relevant.  Following the $N$-body simulation work, we will assume either NFW or Einasto for the parametric form of the profile. As for its exact parameter values, one possibility would be to adopt those given by state-of-the-art $N$-body simulations of MW-size halos, such as Via Lactea II \cite{Diemand:2008in} for the NFW profile or Aquarius \cite{Springel:2008cc} for Einasto. However, although extremely useful to understand what would be {\it typically} expected for MW-like halos, these simulations may provide values of the relevant profile parameters that could significantly differ from the actual ones for the MW. For this reason, and because there is much more data available for the MW than for any other galaxy, it would be desirable to base our specific profile parameter choices on observations, even if the current uncertainties are large. We will do so by following the recent work in Ref.~\cite{Pato:2015dua}, where the authors performed the most complete and up-to-date compilation of astronomical kinematic tracers at different Galactocentric distances, and used them to set dynamical constraints on the MW DM  density profile. By fitting all available data to NFW and Einasto, they inferred the favored ranges of profile parameters for each of these two cases. We adopt the best-fit values in Ref.~\cite{Pato:2015dua}. These correspond to $\rho_{local}=0.42$~GeV~cm$^{-3}$ for the local DM density at the Solar Galactocentric radius ($r_\odot = 8\,$kpc), for both NFW and Einasto, and $\alpha = 0.36$ for the Einasto parameter in Eq.~\eqref{eq:einasto}. 
The results of  Ref.~\cite{Pato:2015dua} are obtained for a scale radius value of $r_s=20\,$kpc, and are not very sensitive to the variations of the latter.
 We follow Ref.~\cite{Catena:2009mf} 
 --- also based on observational data and including dynamical constraints at $\sim$ 20--100\,kpc Galactocentric distances --- and also
  adopt $r_s = 20\,$kpc for the two DM density profiles considered. 
 Hence, in summary, we use $\rho_{local}=0.42$~GeV~cm$^{-3}$, $r_\odot = 8\,$kpc, $\alpha = 0.36$, and $r_s = 20\,$kpc.

\begin{figure}[thb]
\centering
\includegraphics[width = 0.9\linewidth]{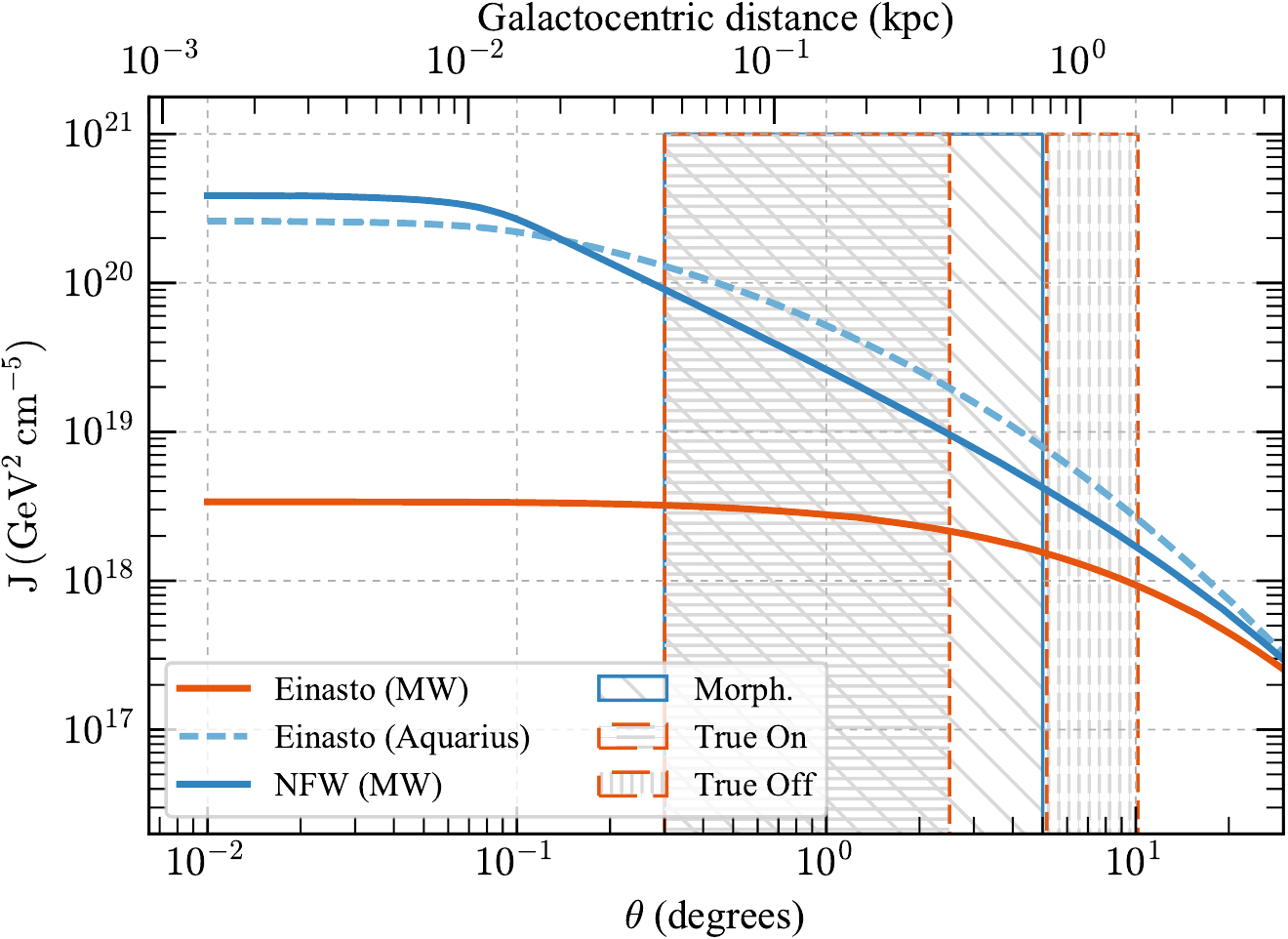}
\caption{\label{fig:J} J~factor as a function of angular distance from the GC for the two profiles considered in this work 
(derived from a fit to the rotation curve of the Milky Way \cite{Pato:2015dua})
with those parameters in the text (Sec.~\ref{sec:profiles}): NFW (solid blue line) and Einasto (solid orange). For comparison, we also show as a dashed blue line the J~factor given by the Einasto profile used in Refs.~\cite{Silverwood:2014yza,HESS:2016jja,Carr:2015hta}.
For each value of $\theta$ the integral over the solid angle in Eq.~\eqref{eq:annflux} is evaluated over an azimuthal angle interval of $0.1^\circ$.
Regions of the GC halo that will be used in our analysis are also depicted in the figure as striped areas delimited by vertical lines (see legend). These regions depend on the observational strategy adopted in each case, see Sec. \ref{sec:obs_strategy} for full details.}
\end{figure}

We note that other parameter choices have been made for the DM density profiles in previous work. For instance, the authors of Ref.~\cite{Silverwood:2014yza} use an
Einasto profile with $\rho_{local} = 0.4\unit{GeV}\unit{cm}^{-3}$ at $r_\odot = 8.5\,$kpc, $\alpha = 0.17$, and $r_s = 20\,$kpc. The same values were also adopted in the recent analysis of the GC halo by the H.E.S.S. Collaboration \cite{HESS:2016jja}, and are partially motivated by the results of the Aquarius $N$-body simulations \cite{Springel:2008cc,Pieri:2009je}.

With these parameters at hand we use the CLUMPY code \cite{2012CoPhC.183..656C,Bonnivard:2015pia} to calculate the J~factor for both NFW and Einasto profiles with the parameters given above. More specifically, CLUMPY provides all-sky J~factor maps for each DM profile in Galactic coordinates.
The resulting J~factors as a function of angular distance from the GC can be seen in Fig.~\ref{fig:J}. For the sake of comparison, we also show in the same figure the J~factor given by the Einasto profile used in Refs.~\cite{Silverwood:2014yza,HESS:2016jja,Carr:2015hta}. Note that the latter profile yields a J~factor which is indeed more similar to the one obtained with our NFW profile rather than with the Einasto profile we use. 

Finally, we note that only the smooth DM component of the Galaxy has been included in either case, as any possible enhancement due to halo substructure is expected to be very marginal in the inner Galactic regions, where these analyses are performed  \cite{SanchezConde:2011ap,2010JCAP...07..023P}. 

Having discussed the astrophysical inputs relevant for our reasoning we will describe the particle physics interactions we are probing. We start with the EFT framework.


\subsection{Effective field theory}   \label{sec:EFTtheory}

We examine four dimension-six benchmark operators describing Dirac fermion DM interacting with SM quarks via scalar, pseudoscalar, vector, and axial-vector interactions~\cite{Goodman:2010ku},
\begin{eqnarray}
\mathcal{O}_S &=& \frac{m_q}{M_\star^3}(\bar\chi\chi )(\bar q q), \\
\mathcal{O}_P &=& \frac{m_q}{M_\star^3}(\bar\chi\gamma^5\chi) (\bar q\gamma^5 q),\\
\mathcal{O}_V &=& \frac{1}{M_\star^2}(\bar\chi\gamma^\mu\chi) (\bar q\gamma_\mu q),\\
\mathcal{O}_A &=& \frac{1}{M_\star^2}(\bar\chi\gamma^\mu\gamma^5\chi) (\bar q\gamma_\mu \gamma^5 q).
\end{eqnarray}
Here $M_\star$ is the energy scale describing the strength of the interaction, and $\gamma_\mu, \, \gamma_5$ are the standard Dirac gamma matrices. These operators were chosen since they display various types of suppression of the annihilation and scattering rate, summarized in Table~\ref{eftSuppression}. 
The annihilation rate is $p$-wave suppressed for operator $\mathcal{O}_S$, and so is proportional to the DM velocity squared $v^2 \sim 10^{-6}$. Operator  $\mathcal{O}_A$ has a $p$-wave suppressed term, and a helicity-suppressed $s$-wave term proportional to $m_q^2$. 
Therefore we expect ID constraints to be relatively weaker for these operators.
For operators $\mathcal{O}_P$ and $\mathcal{O}_A$, 
the scattering rates are either suppressed by the spin of the target nucleus $\vec{s}_N$ or the scattering momentum exchange $\vec{q}$ or both, rendering weak DD constraints.
Operators $\mathcal{O}_S$ and $\mathcal{O}_P$ have interaction strengths suppressed by a Yukawa coupling in order to be consistent with the principle of minimal flavor violation \cite{D'Ambrosio:2002ex,Goodman:2010ku,Goodman:2010qn,Abdallah:2015ter}. This suppresses the ID rate especially when annihilation to top quarks is not kinematically accessible.
It also leads to relatively weaker collider constraints.
The operator $\mathcal{O}_V$ is unique amongst our choice of operators in that it has an unsuppressed rate for collider, DD and ID experiments.
To ease comparison with collider constraints, we assume that the DM couples only to quarks, with an equal coupling to each generation; i.e. $M_\star$ is independent of flavor. 
\begin{table}[t]
\caption{\label{tab:suppression}Summary of the suppression effects the four operators lead to in indirect and direct detection of DM.}
\begin{center}
\begin{tabular}{c|cc}
\hline
\hline
     &  ID   &   DD \\ \hline
$\mathcal{O}_S$   &  $v^2$ & 1 \\ 
$\mathcal{O}_P$   &  1    & $(\vec{s}_\chi \cdot \vec{q}) \,(\vec{s}_N \cdot \vec{q})$ \\ 
$\mathcal{O}_V$   &  1    & 1 \\ 
$\mathcal{O}_A$   &  $m_q^2$, $v^2$  & $\vec{s}_\chi \cdot \vec{s}_N$ \\
\hline
\end{tabular}
\end{center}
\label{eftSuppression}
\end{table}%

\begin{figure*}[thb]
\centering
\includegraphics[width = 0.65\linewidth]{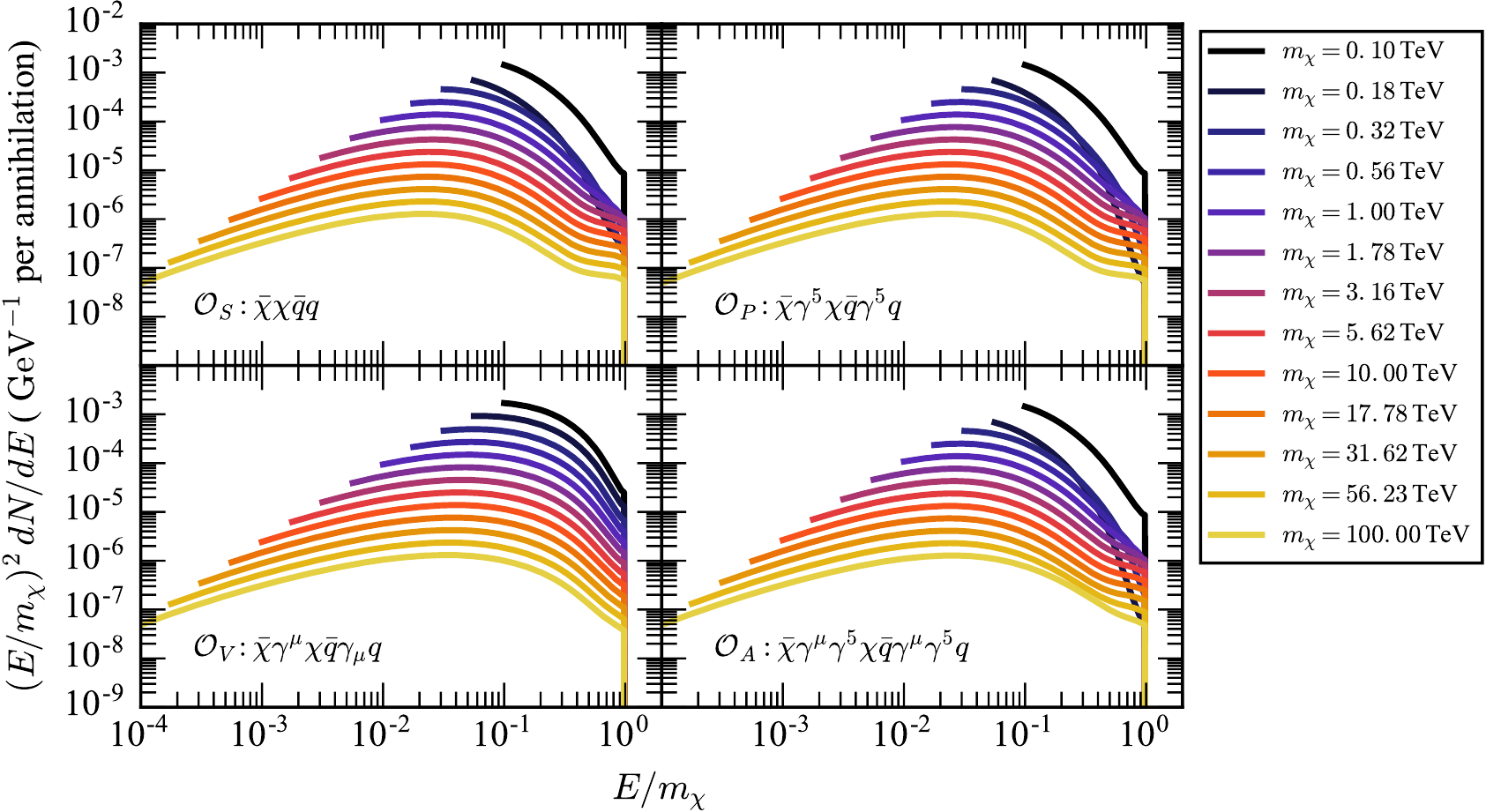}
\caption{\label{fig:specEFT} Average photon spectrum per DM annihilation into quarks. Photon spectra are from \textsc{PPPC4DMID} \cite{Cirelli:2010xx,Ciafaloni:2010ti} and branching ratios are determined using Eqs.~(\ref{sigvOS})--(\ref{sigvOA}).}
\end{figure*}

We use the \textsc{PPPC4DMID} code \cite{Cirelli:2010xx,Ciafaloni:2010ti} to determine the spectrum of photons induced by annihilation into quarks. This is the spectrum at source, and includes the effects of electroweak radiative corrections but neglects secondary photons produced during propagation to Earth such as from inverse Compton scattering and synchrotron emission. The branching ratios and the conversion between limits on $\langle\sigma v\rangle$ and $M_\star$ are given by the DM annihilation rates for each operator,
\begin{eqnarray}
\langle\sigma v\rangle_{\mathcal{O}_S} &=& \sum_q \Theta(m_\chi - m_q)\frac{m_q^2}{M_\star^6}\frac{3 m_\chi^2}{8 \pi} \left(1-\frac{m_q^2}{m_\chi^2}\right)^{3/2} \hspace{-5pt}v^2,\label{sigvOS}\\
\langle\sigma v\rangle_{\mathcal{O}_P} &=& \sum_q\Theta(m_\chi - m_q)\frac{m_q^2}{M_\star^6} 
\frac{3 m_\chi^2}{16\pi} \sqrt{1-\frac{m_q^2}{m_\chi^2}} \nonumber\\
&{}&\times\left(8+\frac{2-m_q^2/m_\chi^2}{1-m_q^2/m_\chi^2}v^2\right),\label{sigvOP}\\
\langle\sigma v\rangle_{\mathcal{O}_V} &=& \sum_q\Theta(m_\chi - m_q)\frac{1}{M_\star^4}\frac{m_\chi^2}{2\pi} \sqrt{1-\frac{m_q^2}{m_\chi^2}}\nonumber\\
&{}&\times\left( 6+3\frac{m_q^2}{m_\chi^2} +\frac{8-4 {m_q^2}/{m_\chi^2}+5 {m_q^4}/{m_\chi^4}}{8  \left(1-{m_q^2}/{m_\chi^2}\right)}v^2\right),\nonumber\\
&{}&\label{sigvOV}\\
\langle\sigma v\rangle_{\mathcal{O}_A} &=& \sum_q\Theta(m_\chi - m_q)\frac{1}{M_\star^4}\frac{m_\chi^2}{4 \pi}\sqrt{1-\frac{m_q^2}{m_\chi^2}}\nonumber\\
&{}&\times\left(6\frac{m_q^2}{m_\chi^2} +\frac{8-22 {m_q^2}/{m_\chi^2}+17 {m_q^4}/{m_\chi^4}}{4 \left(1-{m_q^2}/{m_\chi^2}\right)}v^2\right),\nonumber\\
&{}&\label{sigvOA}
\end{eqnarray}
where $\Theta$ is the Heaviside function, enforcing that DM can only annihilate to kinematically accessible states. 
Using the Heavyside function implies that we only take on-shell two-body final states into account. Allowing off-shell production would smooth the step functions and could significantly change the branching ratios near the threshold. 
The resultant spectra are shown in Fig.~\ref{fig:specEFT}.
For all operators but $\mathcal{O}_V$, 
one can see a jump in the hardness in the annihilation spectra
where annihilation into $t\bar{t}$ quarks is kinematically accessible.
It arises because these three models have leading terms in the annihilation proportional to the quark mass.
For the same reason, 
the spectra for these operators and DM masses 
are very similar when represented as $dN / dx$ versus $x$, where $x = E/m_\chi$. 

\textsc{PPPC4DMID} accounts for electroweak corrections up to next-to-leading-order (NLO) level. For annihilation into quarks this is 
done through the results of Ref.~\cite{Ciafaloni:2010ti}. 
The NLO approximation breaks down for DM masses beyond 10\,TeV and therefore we will not present limits 
on the EFT scale for larger values of $m_\chi$.

As aforementioned the EFT breaks down in some regimes, and for this reason, simplified models have become powerful tools to explore DM models, as we discuss below.

\subsection{Simplified models}  \label{sec:simpmods}

In order to aid in comparison with constraints from other experiments, we study four models  recommended by the LHC Dark Matter Working Group (DMWG) \cite{Boveia:2016mrp},
with Dirac fermion DM interacting with SM quarks via $s$-channel exchange of a scalar, pseudoscalar, vector, or axial-vector mediator (referred to as the $S$, $P$, $V$, and $A$ models in the following).  They correspond to the simplest realization of the four effective operators we consider, and again nicely demonstrate the different classes of suppression. These models are described in detail in Ref.~\cite{Boveia:2016mrp}, and have interaction Lagrangians
\begin{eqnarray}
\mathcal{L}_S &=& -g_\chi \phi \bar \chi \chi - \sum_q g_q \phi \frac{m_q}{\vev} \bar q q, \\
\mathcal{L}_P &=& -i g_\chi \phi \bar \chi \gamma_5 \chi - \sum_q i g_q \phi \frac{m_q}{\vev} \bar q \gamma_5 q, \\
\mathcal{L}_V &=& -g_\chi  Z'_\mu \bar \chi \gamma^\mu \chi - \sum_q g_q Z'_\mu \bar q\gamma^\mu q, \\
\mathcal{L}_A &=& -g_\chi Z'_\mu \bar \chi \gamma^\mu \gamma_5 \chi - \sum_q g_q Z'_\mu \bar q \gamma^\mu \gamma_5 q,
\end{eqnarray}
where $\vev \simeq 246$ GeV is the Higgs vacuum expectation value, the factor of $m_q / \vev$ comes from a Yukawa coupling, and we assume $g_q$ is the same for all quarks.
We follow the DMWG benchmark coupling strengths of $g_q = g_\chi = 1$ for the $S$ and $P$ models, $g_q = 0.25$, $g_\chi = 1$ for the $V$ and $A$ models, and present exclusion contours in the $M_{\rm med}$-$m_\chi$ plane.

When kinematically accessible, each of these mediators has a decay width into fermions (quarks or DM) given by
\begin{eqnarray}
\Gamma_{S}^{f\bar f} &=& c_f^{SP}  \frac{g_f^2 M_{\rm med}}{8 \pi} \left(1-\frac{4m_f^2}{M_{\rm med}^2}\right)^{3/2}, \label{widthS}\\
\Gamma_{P}^{f\bar f} &=& c_f^{SP}  \frac{g_f^2 M_{\rm med}}{8 \pi} \sqrt{1-\frac{4m_f^2}{M_{\rm med}^2}},  \\
\Gamma_{V}^{f\bar f} &=& c_f^{VA}  \frac{g_f^2 M_{\rm med}}{12 \pi} \sqrt{1-\frac{4m_f^2}{M_{\rm med}^2}} \left(1+\frac{2m_f^2}{M_{\rm med}^2}\right),  \\
\Gamma_{A}^{f\bar f} &=&  c_f^{VA}  \frac{g_f^2 M_{\rm med}}{12 \pi} \left(1-\frac{4m_f^2}{M_{\rm med}^2}\right)^{3/2} ,\label{widthA}
\end{eqnarray}
where $c_\chi^{SP}=c_\chi^{VA} = 1$, $c_q^{SP} = 3 m_q^2 / \vev^2$, and $c_q^{VA} = 3$. In the $S, P$ models the mediator also decays to gluons,
\begin{eqnarray}
\Gamma_{S}^{gg} &=& \frac{ \alpha_S^2 g_q^2 M_{\rm med}^3 \left| f_{\rm scalar}\left(\frac{4 m_t^2}{M_{\rm med}^2}\right)\right| ^2}{32 \pi ^3 \vev^2},\\
\Gamma_{P}^{gg} &=& \frac{ \alpha_S^2 g_q^2 M_{\rm med}^3 \left| f_{\rm pseudoscalar}\left(\frac{4 m_t^2}{M_{\rm med}^2}\right)\right| ^2}{32 \pi ^3 \vev^2}, \label{widthPgg}
\end{eqnarray}
where
{
\allowdisplaybreaks
\begin{eqnarray}
f_{\rm scalar}(\tau) = \tau  \left(1+(1-\tau ) \tan^{-1}\left(\frac{1}{\sqrt{\tau -1}}\right)^2\right) ,\\
f_{\rm pseudoscalar}(\tau) = \tau \tan^{-1}\left(\frac{1}{\sqrt{\tau -1}}\right)^2.
\end{eqnarray}
}
The total minimum decay width  is then 
{
\allowdisplaybreaks
\begin{eqnarray}
\Gamma_{S,\rm tot} &=& \Gamma_S^{gg} + \sum_q \Theta(M_{\rm med} - 2m_q) \Gamma_S^{q \bar q}\nonumber\\
&{}&+\Theta(M_{\rm med} - 2m_\chi)\Gamma_S^{\chi \bar \chi},\\
\Gamma_{P,\rm tot} &=& \Gamma_P^{gg} + \sum_q \Theta(M_{\rm med} - 2m_q) \Gamma_P^{q \bar q}\nonumber\\
&{}&+  \Theta(M_{\rm med} - 2m_\chi)\Gamma_P^{\chi \bar \chi},\\
\Gamma_{ V,\rm tot} &=&\sum_q \Theta(M_{\rm med} - 2m_q) \Gamma_V^{q \bar q}\Theta(M_{\rm med} - 2m_\chi)\Gamma_V^{\chi \bar \chi},\nonumber\\
\\
\Gamma_{A,\rm tot} &=&\sum_q \Theta(M_{\rm med} - 2m_q) \Gamma_A^{q \bar q} \Theta(M_{\rm med} - 2m_\chi)\Gamma_A^{\chi \bar \chi}\nonumber\\
\end{eqnarray}
}
%
In principle the mediator could couple to other final states such as leptons, though we assume only the minimum width. 
These simplified models have similar DM annihilation rates to quarks as the effective operators described earlier,
%
{
\small
\allowdisplaybreaks
\begin{eqnarray}
\langle\sigma v\rangle_{S}^{q \bar q} &=& \frac{3 g_q^2 g_\chi^2}{8 \pi} \frac{m_q^2}{\vev^2} \frac{m_\chi^2 \left(1-\frac{m_q^2}{m_\chi^2}\right)^{3/2}}{\left(M_{\rm med}^2-4 m_\chi^2\right)^2+\Gamma_{S,\rm tot} ^2 M_{\rm med}^2}v^2,\label{sigvqqS}\\
 \langle\sigma v\rangle_{P}^{q \bar q} &=& \frac{3 g_q^2 g_\chi^2}{2 \pi} \frac{m_q^2}{\vev^2}  \frac{m_\chi^2 \sqrt{1-\frac{m_q^2}{m_\chi^2}}}{\left(M_{\rm med}^2-4 m_\chi^2\right)^2+\Gamma_{P, \rm tot} ^2 M_{\rm med}^2},\nonumber\\
 &{}& +\mathcal{O}(v^2 )\label{sigvqqP}\\
\langle\sigma v\rangle_{V}^{q \bar q} &=&  \frac{3 g_q^2 g_\chi^2}{2\pi} \frac{ \left(m_q^2+2 m_\chi^2\right)\sqrt{1-\frac{m_q^2}{m_\chi^2}}}{ \left(M_{\rm med}^2-4 m_\chi^2\right)^2+\Gamma_{V,\rm tot} ^2 M_{\rm med}^2}+\mathcal{O}(v^2 ),\label{sigvqqV}\\
\langle\sigma v\rangle_{A}^{q \bar q} &=& \frac{g_q^2 g_\chi^2}{2\pi}   \nonumber\\
&{}&\times\frac{3m_q^2 \left(1-\frac{4 m_\chi^2}{M_{\rm med}^2}\right)^2 \sqrt{1-\frac{m_q^2}{m_\chi^2}}+m_\chi^2 v^2+\mathcal{O}(m_q^2 v^2 )}{\left(M_{\rm med}^2-4 m_\chi^2\right)^2+\Gamma_{A,\rm tot} ^2 M_{\rm med}^2}\nonumber\\
&{}&+\mathcal{O}(v^4 ) \label{sigvqqA}.
\end{eqnarray}
}
%
In the $S$ and $P$ models, DM can also annihilate to gluons via a quark loop. Across most of the parameter space this is subdominant to direct annihilation into quarks or the mediator, but can be important when the DM is light and annihilation to top quarks or mediators is not kinematically allowed, and is therefore included for completeness. The annihilation rate is given by
%
{
\allowdisplaybreaks
\begin{eqnarray}
\langle\sigma v\rangle_{S}^{gg} &=&\frac{\alpha^2}{8 \pi^3 \vev^2}%
\frac{2 g_q^2 g_\chi^2 m_\chi^4 v^2}{\left(M_{\rm med}^2-4 m_\chi^2\right)^2+\Gamma_{S,\rm tot} ^2 M_{\rm med}^2}\nonumber\\
&{}&\times\left|\sum_q f_{\rm scalar}\left(\frac{m_q^2}{m_\chi^2}\right)\right| ^2  ,\label{sigvggS}\\
\langle\sigma v\rangle_{P}^{gg} &=&\frac{\alpha^2}{2 \pi^3 \vev^2}%
\frac{g_q^2 g_\chi^2}{\left(M_{\rm med}^2-4 m_\chi^2\right)^2+\Gamma_{P,\rm tot} ^2 M_{\rm med}^2}\nonumber\\
&{}&\times\left|\sum_q m_q^2 f_{\rm pseudoscalar}\left(\frac{m_q^2}{m_\chi^2}\right)\right| ^2 \label{sigvggP}. 
\end{eqnarray}
}
%
When the DM mass is heavier than the mediator mass, direct annihilation to the mediator becomes accessible, with the mediator subsequently decaying into SM particles. The annihilation rates are 
%

{
\allowdisplaybreaks
\begin{eqnarray}
\langle\sigma v\rangle_{S}^{SS} &=& \frac{g_\chi^4 m_\chi v^2}{24 \pi  }\nonumber\\
&{}&\times\frac{\left(9 m_\chi^6-17 M_{\rm med}^2 m_\chi^4+10 M_{\rm med}^4 m_\chi^2-2 M_{\rm med}^6\right)}{\sqrt{m_\chi^2-M_{\rm med}^2} \left(2 m_\chi^2-M_{\rm med}^2\right)^4},\nonumber\\
&{}&\label{sigvSS}\\
\langle\sigma v\rangle_{P}^{PP} &=&\frac{g_\chi^4 m_\chi \left(m_\chi^2-M_{\rm med}^2\right)^{5/2}}{24 \pi  \left(2m_\chi^2-M_{\rm med}^2\right)^4}v^2 ,\label{sigvPP}\\
\langle\sigma v\rangle_{V}^{VV} &=&\frac{g_\chi^4 \left(m_\chi^2-M_{\rm med}^2\right)^{3/2}}{4 \pi  m_\chi \left(M_{\rm med}^2-2 m_\chi^2\right)^2}+\mathcal{O}(v^2 ),\label{sigvVV}\\
\langle\sigma v\rangle_{A}^{AA} &=&\frac{g_\chi^4 \left(m_\chi^2-M_{\rm med}^2\right)^{3/2}}{4 \pi  m_\chi \left(M_{\rm med}^2-2 m_\chi^2\right)^2}+\mathcal{O}(v^2 ) \label{sigvAA}.
\end{eqnarray}
}

The total annihilation cross sections are then:
{
\allowdisplaybreaks
\begin{eqnarray}
\langle\sigma v\rangle_{S}^\mathrm{total} &=& \langle\sigma v\rangle_{S}^{gg} +  \Theta(m_\chi - M_{\rm med}) \langle\sigma v\rangle_{S}^{SS} \nonumber\\
&{}&+ \sum_q \Theta(m_\chi - m_q) \langle\sigma v\rangle_{S}^{q \bar q},\label{sigvtotS}\\
\langle\sigma v\rangle_{P}^\mathrm{total} &=& \langle\sigma v\rangle_{P}^{gg} +  \Theta(m_\chi - M_{\rm med}) \langle\sigma v\rangle_{P}^{PP} \nonumber\\
&{}&+ \sum_q \Theta(m_\chi - m_q) \langle\sigma v\rangle_{P}^{q \bar q},\label{sigvtotP}\\
\langle\sigma v\rangle_{V}^\mathrm{total} &=& \Theta(m_\chi - M_{\rm med}) \langle\sigma v\rangle_{V}^{VV} \nonumber\\
&{}&+ \sum_q \Theta(m_\chi - m_q) \langle\sigma v\rangle_{V}^{q \bar q},\label{sigvtotV}\\
\langle\sigma v\rangle_{A}^\mathrm{total} &=& \Theta(m_\chi - M_{\rm med}) \langle\sigma v\rangle_{A}^{AA} \nonumber\\
&{}&+ \sum_q \Theta(m_\chi - m_q) \langle\sigma v\rangle_{A}^{q \bar q}.\label{sigvtotA}
\end{eqnarray}
}

\begin{figure*}[thb]
\centering
\includegraphics[width = 0.75\linewidth]{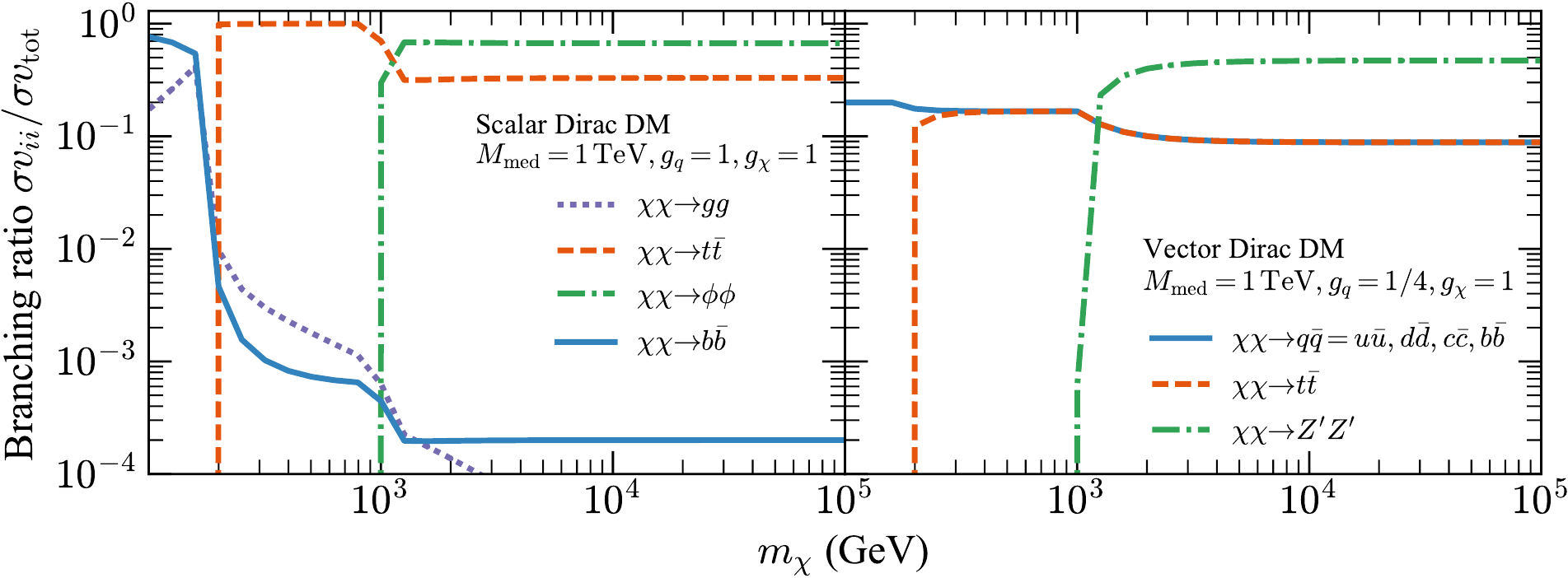}
\caption{\label{fig:SVBr} Branching ratios for the $S$ and $V$ models for $g_q = g_\chi = 1$ ($S$) and $g_q =0.25, g_\chi = 1$ ($V$), with $M_\mathrm{med} = 1\,$TeV.}
\end{figure*}

The branching ratios to various final states are shown in Fig.~\ref{fig:SVBr} for the $S$ and $V$ models. 
For the $A$ model, annihilation to the heaviest kinematically accessible quark dominates. For the $P$ model, annihilation to top quarks dominates when kinematically accessible. Below this threshold, gluons are the leading annihilation channel.

\begin{figure*}[thb]
\centering
\includegraphics[width = 0.75\linewidth]{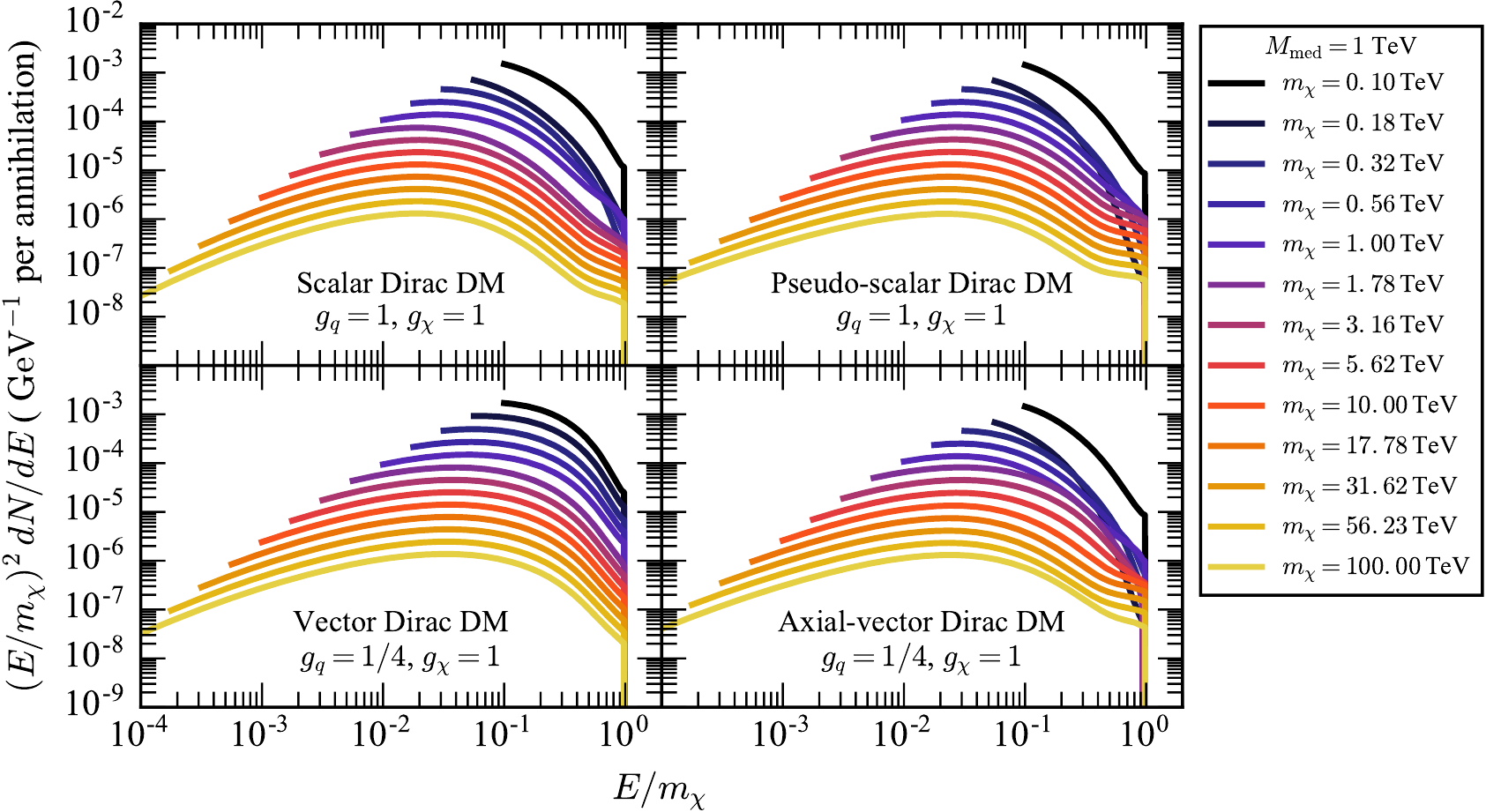}
\caption{\label{fig:specSM} Simplified model \textsc{PPPC4DMID} spectra for $M_\mathrm{med} = 1\,$TeV}
\end{figure*}

The photon spectra per DM annihilation to quarks and gluons are again determined using \textsc{PPPC4DMID}.
The photon spectrum per annihilation into mediators is a little more involved. The spectrum from the decay of a mediator into quarks and gluons is calculated using \textsc{PPPC4DMID} in the mediator rest frame as usual, with branching ratios from Eqs.~(\ref{widthS})--(\ref{widthPgg}). These spectra are then Lorentz boosted into the DM center-of-mass frame using the procedure from Ref.~\cite{Bergstrom:2008ag,Elor:2015tva}. For each model, the spectra from annihilation to quarks, gluons and mediators are combined and weighted by their respective branching ratios using Eqs.~(\ref{sigvqqS})--(\ref{sigvAA}). Results are shown in Fig.~\ref{fig:specSM}.
The spectra are very similar to the EFT case, in the sense that 
jumps in the spectral hardness are observed 
once annihilation into $t\bar{t}$ becomes kinematically possible for 
the $S$, $P$, and $A$ models. 
In this figure we can also see the resonant enhancement of the annihilation rate around the region $M_{\rm med} \simeq 2m_\chi$. 

After discussing the important ingredients for the determination of the expected $\gamma$-ray signal from DM annihilation, we describe the procedure used to assess CTA sensitivity and the expected $\gamma$-ray backgrounds.

\section{Expected Backgrounds}
\label{sec:bkg}
For IACTs the major source of background 
is cosmic rays (CRs), which consist mainly of protons but also heavier nuclei,
as well as electrons and positrons. 
The flux of CRs is in general by a factor of $10^3$ larger than \gray signals from point sources, requiring efficient techniques to reject showers initiated by CRs (see e.g. Ref.~\cite{voelk2009}).
This can be achieved
 by means of the shower image, 
and potentially the arrival time of the shower front~\cite{Aharonian:2008review}.
However, a residual contamination 
of the $\gamma$-ray sample with CRs is inevitable. 
The expected CR background for CTA 
has been derived through extensive Monte Carlo simulations taking 
into account the different possible array layouts
\cite{Bernlohr:2012we}.\footnote{See also \url{https://www.cta-observatory.org/science/cta-performance/}}
Here, we use the so-called \texttt{prod\,2} version of these background simulations
to derive the expected number of CR background events. 
Due to the soft CR spectrum (see e.g. Ref.~\cite[]{Olive:2016xmw} for a review),
this background component will be especially important at low 
energies, but we expect it to dominate over the entire energy range 
in comparison to a DM signal with an annihilation 
cross section yielding the expected DM relic abundance, 
$\langle \sigma v \rangle \sim 3\times10^{-26}\,\mathrm{cm}^{3}\,\mathrm{s}^{-1}$.

An additional source of background is astrophysical Galactic diffuse emission (GDE) 
caused by the interaction of CR with interstellar dust and radiation fields. 
Below 100\,GeV, the GDE has been measured with the \FermiLAT and found to be  dominated by $\pi^0$ decay, inverse Compton scattering, as well as bremsstrahlung, and the first two contributions dominate for energies above a few GeV \cite{Ackermann:2012pya}.
At energies between 0.2 and 20\,TeV, diffuse \gray emission has been detected with H.E.S.S. from the GC ridge for Galactic latitudes $|b|<0.3^\circ$ and $|l| < 0.8^\circ$ \cite{Aharonian:2006au}.
The authors of Ref. \cite{Silverwood:2014yza} found that 
neglecting GDE leads to a strong overestimation 
of the differential sensitivity for the DM signal from the GC.
We therefore follow Ref. \cite{Silverwood:2014yza} and estimate 
the background contribution of the GDE with the template provided by the \fermiLAT Collaboration.\footnote{\url{http://fermi.gsfc.nasa.gov/ssc/data/access/lat/BackgroundModels.html}}
We use a simple power-law extrapolation of the template for \gray energies above 100\,GeV. 
This yields a conservative estimate of the GDE at higher energies
as a potential cutoff in the GDE energy spectrum would yield less background counts. 
In comparison to Ref.~\cite{Aharonian:2006au}, the extrapolation
overestimates the diffuse flux in the same sky region by approximately 2 orders of magnitude. 
For the GDE measurement with Milagro at a median energy of 15\,TeV for latitudes $-2^\circ < b < 2^\circ$ and longitudes $30^\circ < l < 65^\circ$ and $65^\circ < l < 85^\circ$~\cite{Abdo:2008if} the extrapolation overpredicts the flux by more than 4 orders of magnitude.

We neglect any contribution from resolved and unresolved point sources.
One known source in the region is HESS\,1745-303. In a real analysis the source can be simply cut out as done in H.E.S.S. analyses (see e.g. Fig.~1 of the Supplemental Material of Ref.~\cite{Abdallah:2016ygi}). A similar approach could be taken for additional sources identified in the Galactic plane survey which will be conducted with CTA.
Evidence for unresolved sources like millisecond pulsars has been recently found and such a population could explain 
 the \gray excess observed in the Galactic center~\cite{Lee:2015fea,Bartels:2015aea,Fermi-LAT:2017yoi}. If these sources are indeed millisecond pulsars, they should not contribute in the CTA energy range as their spectra usually lexhibit cutoffs at a few tens of GeV.

\section{Analysis framework}
\label{sec:analysis}

We use \textsc{ctools} version 1.0.1 \cite{Knodlseder:2016nnv} to calculate sky maps with the expected 
number of counts from GDE, CR background, and a potential DM signal.\footnote{Specifically, we use the \textsc{ctmodel} tool; see Ref. \cite{Knodlseder:2016nnv} and \url{http://cta.irap.omp.eu/ctools/index.html}.}
The \textsc{ctools} package folds the predicted intensity for the diffuse DM signal and the GDE with the CTA instrumental response functions (IRFs), taking into account the  
point spread function (PSF), which relates the true arrival direction of the $\gamma$ ray $\mathbf{p}$ to the reconstructed direction $\mathbf{p}'$, effective area $A_\mathrm{eff}$, and the energy-dependent size of the field of view (FOV).\footnote{
The differently sized telescopes that cover partly overlapping energy ranges have different FOV making the resulting FOV dependent on energy. 
}
We neglect the energy dispersion, which should 
not have a large effect since all spectral components are smooth 
and do not show narrow features.
The \texttt{prod$\,$2} Monte Carlo CR background templates for the southern CTA baseline array \cite{Bernlohr:2012we} are implemented through 
the CTA IRF background model.\footnote{\url{http://cta.irap.omp.eu/ctools/users/user_manual/getting_started/models.html}}
Within \textsc{ctools}, the PSF, effective area, and background intensity are extrapolated using analytical expressions in order to calculate the off-axis performance.
We calculate the sky maps within six logarithmic energy bins per decade in an energy range 
from 30\,GeV to 100\,TeV and use a pixelization of $0.0625^\circ\,\mathrm{pixel}^{-1}$. 

\subsection{Observational strategy}  \label{sec:obs_strategy}
Within the first three years of CTA operations it is planned to conduct a 
survey of the central Galaxy to achieve a uniform exposure within a $2^\circ$ radius around the GC \cite{Carr:2015hta}. 
As the final layout of the pointing scheme is not yet known, 
we will assume a pointing centered on the GC
and compute the expected number of counts within $10^\circ \times 10^\circ$ sky maps. 
We will refer to this region as FOV. One should keep in mind 
that the FOV is energy dependent.
At low energies, mostly the large-size telescopes will contribute to the sensitivity
which have a field of view of about $4.5^\circ$. At the highest energies, 
the small sized-telescopes contribute most and have a FOV of $\sim9^\circ$.
Therefore,
we do not expect any counts at large angular distances 
from the FOV center at low energies.

DM searches with IACTs are usually performed by dividing the FOV into multiple regions of interest (RoIs), where regions 
with a large expected DM signal are referred to as ``on'' regions whereas regions with
negligible DM contribution are referred to as ``off'' regions.
The off regions provide an estimate for the expected number of background events. 
The sensitivity can be increased by using multiple on and off regions
in order to probe the different spatial morphology of the 
background and DM signal (e.g.~\cite[][]{Silverwood:2014yza,Lefranc:2015pza,HESS:2016jja}). 

\begin{figure*}[thb]
\centering
\includegraphics[width = 0.7\linewidth]{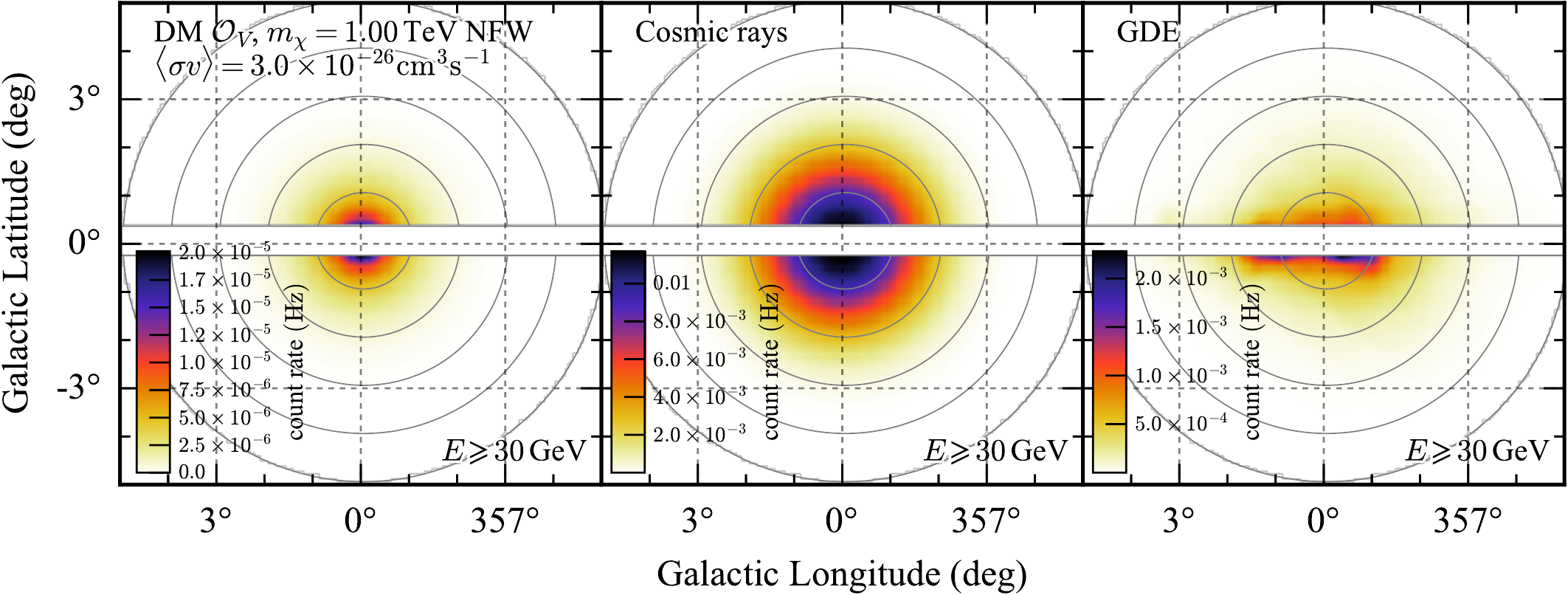}
\includegraphics[width = 0.7\linewidth]{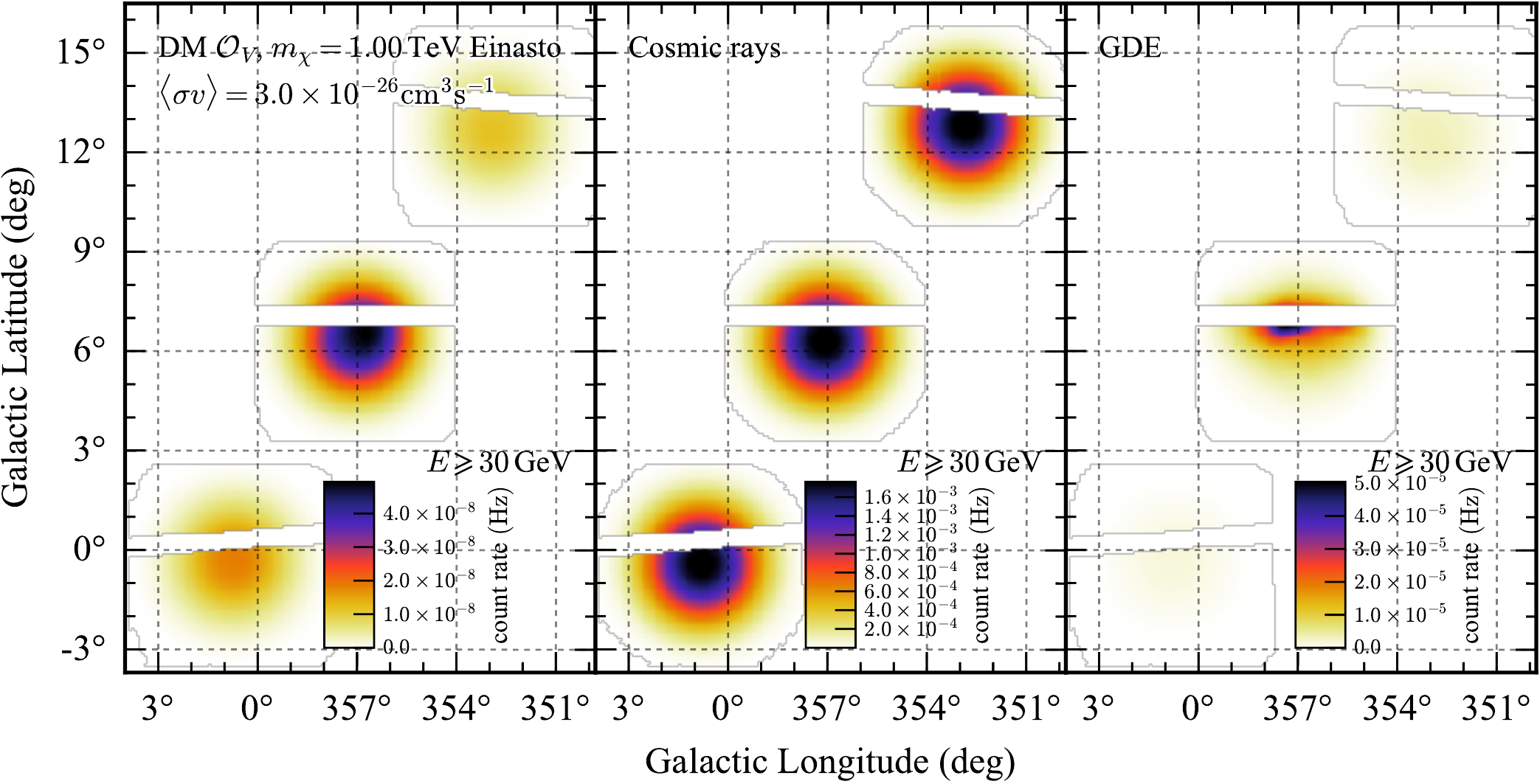}
\caption{\label{fig:morph-point} Adopted pointing schemes.
From left to right the panels show the expected count rate for the
different model components included here (DM, CR, and GDE). 
For the DM, we show the expected rate for the $\mathcal{O}_V$ operator and $m_\chi = 1\,$TeV with $\langle \sigma v \rangle = 3\times 10^{-26}\,\mathrm{cm}^{3}\mathrm{s}^{-1}$.
Top: Pointing schemes for the morphological analysis adopted 
for the NFW profile. Bottom: 
True on/off pointings for the Einasto profile.}
\end{figure*}

For our assumed NFW profile, we follow Refs. \cite{Lefranc:2015pza,Carr:2015hta} and divide the FOV into five concentric rings
with a width of $1^\circ$. 
The outermost ring has an outer radius of $5^\circ$ (see the upper panels of Fig. \ref{fig:morph-point}).
We do not use a separate off region but rather model the contributions from all sources simultaneously \cite{Silverwood:2014yza}.\footnote{
We note that a homogeneous exposure of an inner $2^\circ$ radius will lead to a flatter cosmic-ray spatial profile that will extend to larger distances to the GC as the one we adopt.
}
For our assumed Einasto DM density profile, the FOV will be too small to achieve 
a sufficient contrast between the DM signal and the background with this setup. 
For this profile, we therefore assume ``true'' on/off observations 
with three independent pointings as conducted by the H.E.S.S. Collaboration \cite{HESS:2015cda}.
We use three $5^\circ \times 5^\circ$ RoIs, with the central one centered on $(l,b) = (1^\circ, -0.7^\circ)$ 
and the other two shifted by $\pm 35'$ in right ascension (corresponding to an angular 
separation of $\sim 7.66^\circ$ between the RoI centers).
The RoIs are shown in the lower panels of Fig. \ref{fig:morph-point}.
By consecutively observing the on and off regions, 
differences in azimuth and zenith angle distributions are minimized. 
For all pointing strategies we exclude Galactic latitudes $|b| < 0.3^\circ$ 
to minimize contamination from GDE.
We stress that we do not attempt to optimize the observational strategy
to find the optimal spectral and spatial binning. 
In principle, these should be optimized for different DM density profiles, 
DM spectra, and observation energy (due to the energy dependent FOV). 
This is, however, beyond the scope of this study.

\subsection{Likelihood analysis}
\begin{figure}[thb]
\centering
\includegraphics[width = 0.85\linewidth]{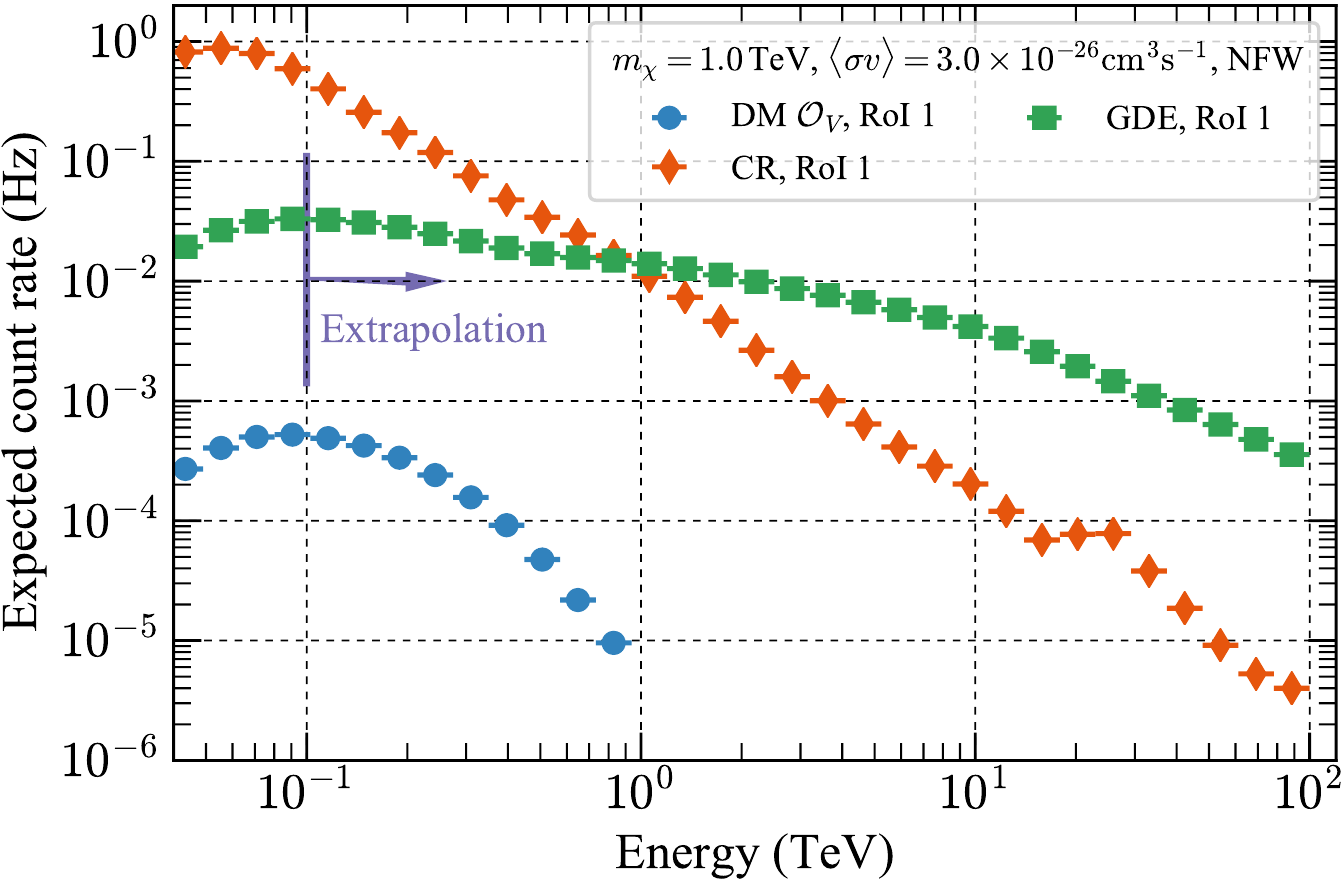}
\includegraphics[width = 0.85\linewidth]{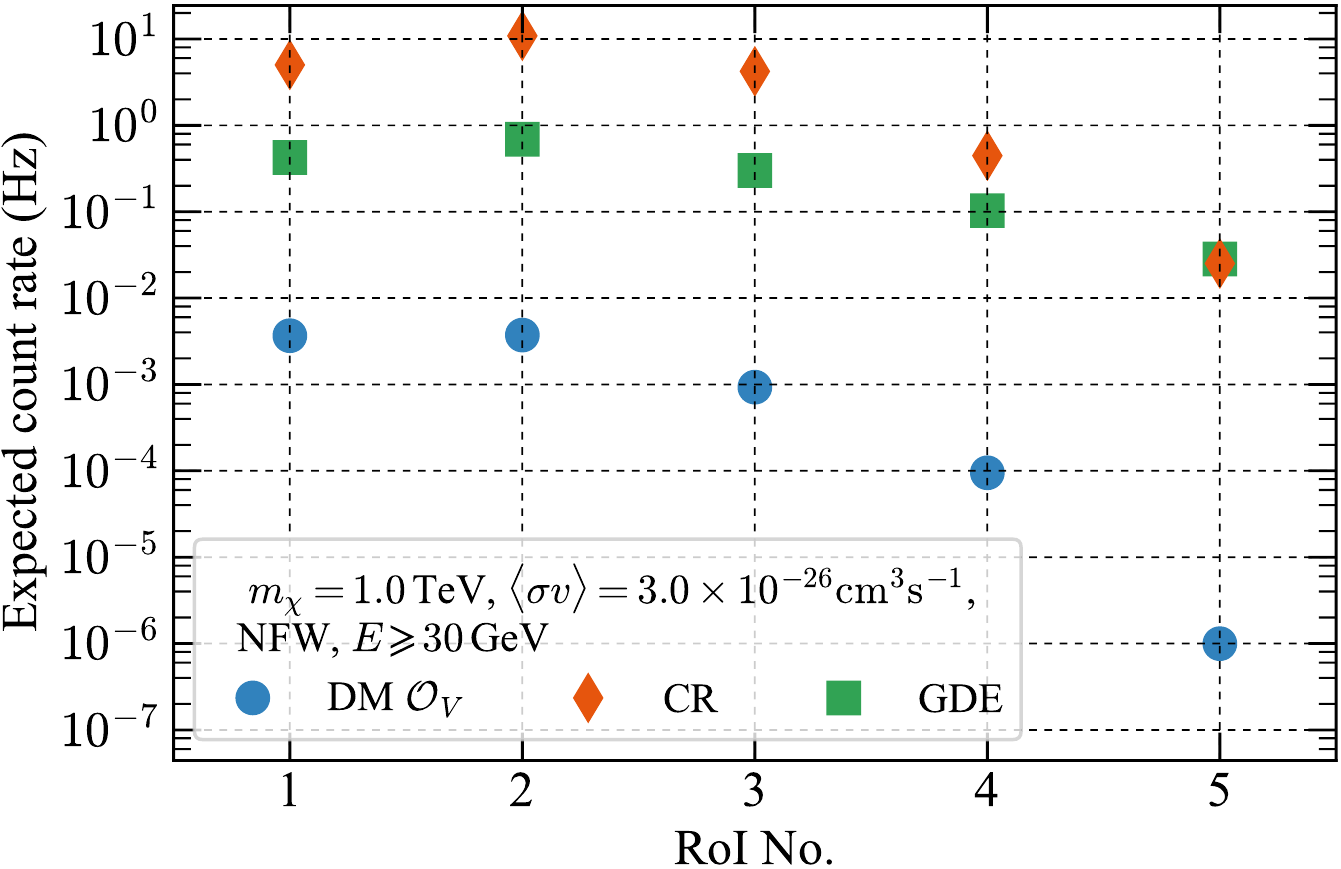}
\caption{\label{fig:counts} Expected count rates for the different source components for the NFW profile and morphological pointing strategy. The DM component is shown for the $\mathcal{O}_V$ operator and $m_\chi = 1\,$TeV with $\langle \sigma v \rangle = 3\times 10^{-26}\,\mathrm{cm}^{3}\mathrm{s}^{-1}$.
Top: Expected count rate as a function of energy 
for the innermost RoI. The energy above which the GDE is extrapolated 
is marked by a blue arrow. 
Bottom: Count rate for each RoI integrated above 30\,GeV.
}
\end{figure}

We use a binned Poisson likelihood analysis 
to derive the CTA sensitivity and closely follow the methodology outlined in Ref. \cite{Silverwood:2014yza}, 
i.e. we do not estimate the background events from independent off regions. 
Instead, we use templates for all model components in all spatial bins and fit each contribution (DM, CR, GDE) simultaneously. 
For the chosen observational strategies we 
use \texttt{ctools} to calculate the expected number 
of counts $\mu_{ik}^\mathrm{X}$ for contribution X$=$DM, GDE, CR, 
for each energy bin $i$ and pixel $k$ within solid angle $\Omega_k$ [see Eq. (4) in Ref.~\cite{Knodlseder:2016nnv}],
\begin{eqnarray}
\mu_{ik}^\mathrm{X} &=& \mathrm{T}_\mathrm{obs}\, \int\limits_{\Omega_k}d\Omega \int\limits_{\Delta E_i} dE \int\limits_{\mathbf{p}}d\mathbf{p}\,
A_\mathrm{eff}(\mathbf{p}, E) \nonumber\\
&{}&\times \mathrm{PSF}(\mathbf{p}' |\mathbf{p}, E)\frac{\mathrm{d}N^\mathrm{X}}{\mathrm{d}E}(E,\mathbf{p}),
\end{eqnarray}
where $\mathrm{T}_\mathrm{obs}$ is the observation time, $\mathbf{p}, \mathbf{p}'$ are the true and reconstructed $\gamma$-ray arrival directions, respectively, and $dN^\mathrm{X}/dE$ is the diffuse model for $\gamma$-ray emission from component X.
The expected counts for all pixels and model components
above 30\,GeV are shown in Fig.~\ref{fig:morph-point}.
The number of expected counts in RoI $j$ is then simply $\mu_{ij}^X=\sum_{k\in\mathrm{RoI}_j} \mu_{ik}^X$ for all pixel $k$ in RoI $j$. 
An example of the resulting count rate spectrum 
for the innermost ring of the pointing strategy adopted for the NFW profile is shown in Fig.~\ref{fig:counts} (top). 
With our chosen extrapolation of the GDE above 100\,GeV, it dominates the count rate above $\sim10\,$TeV. 
The count rate in each ring integrated above 30\,GeV is shown 
in Fig.~\ref{fig:counts} (bottom). 
For a constant acceptance, one would expect 
the CR to increase for the RoIs with larger distances 
to the GC due to the increasing solid angle. 
This is not the case here due to decreasing exposure towards 
the edges of the FOV. 

The total number of expected counts for each energy and RoI is given by the sum 
of the model components:
\begin{equation}
\mu_{ij} = \mu^\mathrm{DM}_{ij} + R^\mathrm{CR}_{i}\mu^\mathrm{CR}_{ij}
 +  R^\mathrm{GDE}_{i}\mu^\mathrm{GDE}_{ij}.
\end{equation}
In the statistical analysis, we allow each component to be rescaled independently in each energy bin. For the DM component this is done by changing $\langle \sigma v \rangle$ while the parameters $R_i^\mathrm{CR,GDE}$ change each background contribution. 
Up to a constant, the likelihood for $\mathbf{n}$ observed number of  counts is given by
\begin{equation}
\ln\mathcal{L}(\boldsymbol{\mu},\boldsymbol{\theta}|\mathbf{n}) = 
\sum\limits_{i,j}\left[n_{ij}\ln\left(\alpha_{ij}\mu_{ij}\right)  - \alpha_{ij}\mu_{ij} - \frac{(1-\alpha_{ij})^2}{2\sigma_\alpha^2}\right], 
\label{eq:lnl}
\end{equation}
where we introduced the terms $\alpha_{ij}$ that allow us to account for systematic uncertainties such as an unmodeled variation in the exposure between the different RoIs. Each $\alpha_{ij}$ is assumed to follow a Gaussian likelihood with a width of $\sigma_\alpha = 0.01$~\cite{Silverwood:2014yza}. The nuisance parameters are given by $\boldsymbol{\theta} = (\mathbf{R}^\mathrm{GDE},\mathbf{R}^\mathrm{CR},\boldsymbol{\alpha})$.\footnote{
In practice, we calculate the likelihood curve $\ln\mathcal{L}_i$ in each energy bin $i$ as a function of $\langle \sigma v \rangle$ where we maximize the likelihood in terms of the nuisance parameters for each 
value of $\langle \sigma v \rangle$. In a second step, given by Eq.~\eqref{eq:lnl}, we sum these curves  over the energy bins, $\sum_i\ln\mathcal{L}_i$, thereby tying $\langle \sigma v \rangle$ over the energy bins. 
}

Instead of simulating a large set of different Poisson realizations 
of the observed counts, we use the ``Asimov data set''; i.e. we set the observed counts 
equal to the number of expected counts, $n_{ij} = \mu_{ij}$ \cite{Cowan:2010js}.
We do not assume any contribution from DM and set 
$\mu_{ij}^\mathrm{DM} = 0$ and $R_i^\mathrm{GDE} = R_i^\mathrm{CR} = 1$. For each tested DM operator and mass, 
we step through $\langle \sigma v \rangle$ and calculate the 
test statistic
\begin{equation}
\lambda(m_\chi, \langle \sigma v \rangle) = 
-2\ln\left(\frac{\mathcal{L}(\boldsymbol{\mu}(m_\chi,\langle \sigma v \rangle),\widehat{\widehat{{\boldsymbol{\theta}}}}(m_\chi,\langle \sigma v \rangle)|\mathbf{n})}{\mathcal{L}(\widehat{\boldsymbol{\mu}},\widehat{\boldsymbol{\theta}}|\mathbf{n})}\right),
\end{equation}
where $\widehat{\widehat{\boldsymbol{\theta}}}(m_\chi,\langle \sigma v \rangle)$ are the nuisance parameters 
that maximize $\mathcal{L}$ for a given set of $m_\chi,\langle \sigma v \rangle$, 
and $\widehat{\boldsymbol{\mu}}$ and $\widehat{\boldsymbol{\theta}}$ 
denote the unconditional maximum likelihood estimators. 
For the simplified models, $\boldsymbol{\mu}$ and $\widehat{\widehat{\boldsymbol{\theta}}}$ (and consequently $\lambda$) additionally depend on $M_\mathrm{med}$.
For each DM operator and 
mass we then set 95\,\% confidence limits on the annihilation cross section that results 
in $\lambda = 2.71$.
Following Ref. \cite{Silverwood:2014yza}, we restrict $0.5 \leqslant R_i^\mathrm{CR} \leqslant 1.5$ and furthermore $0.2 \leqslant R_i^\mathrm{GDE} \leqslant 5$.

\section{Results}
\label{sec:results}
\begin{figure}[thb]
\centering
\includegraphics[width = 0.9\linewidth]{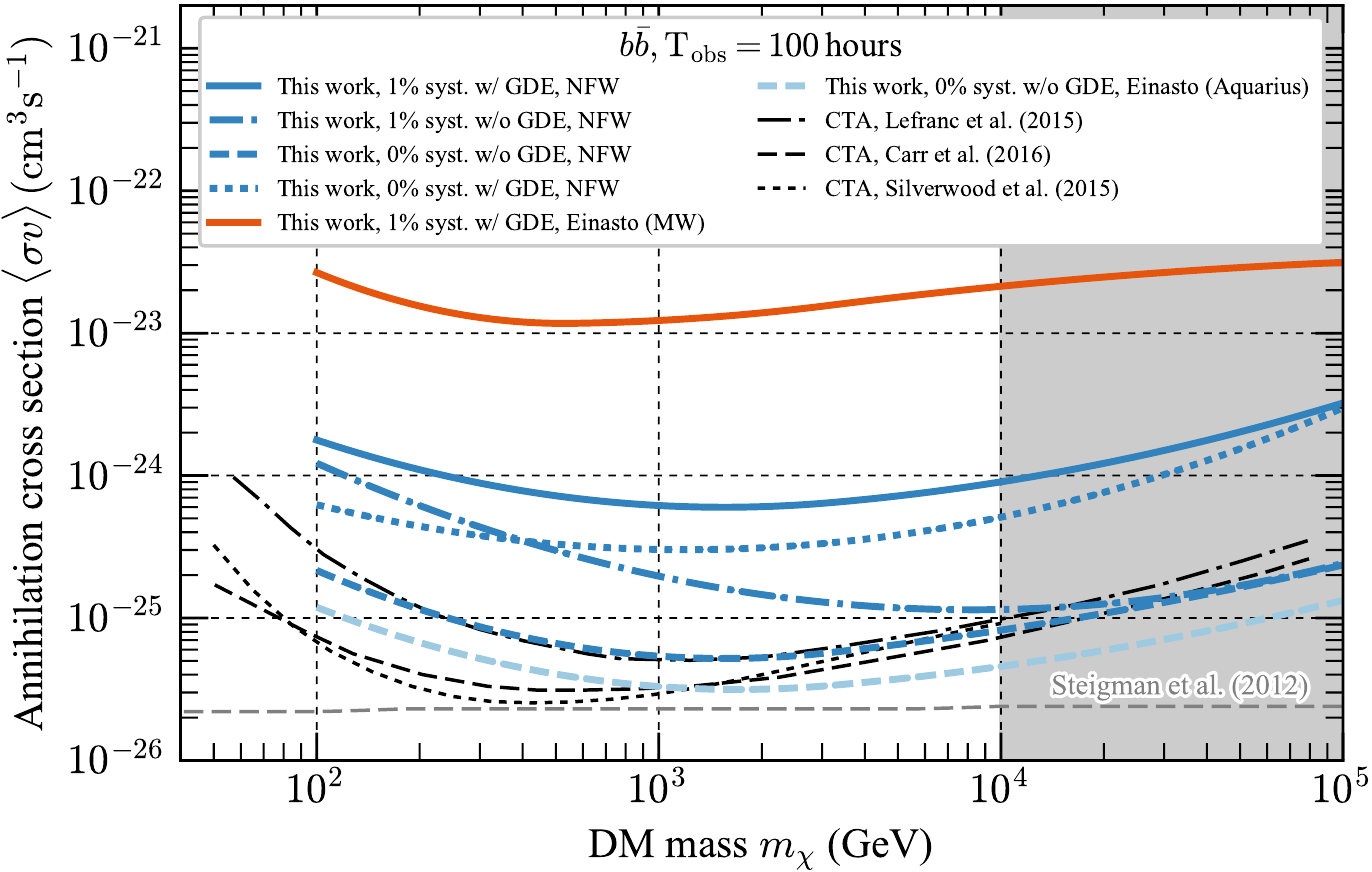}
\caption{\label{fig:bb} Upper limits on the cross section for 100\,\% annihilation of Majorana DM into $b\bar{b}$ for our fiducial analysis (1\,\% systematic uncertainty, including GDE) 
for the two assumed DM density profiles (blue solid line, NFW; orange solid line, Einasto). For comparison we also show our limits for 0\,\% systematic uncertainty for the NFW profile with and without GDE (blue dotted and dashed line, respectively) and for the Einasto profile of the Aquarius simulation (light-blue dashed line). 
The blue dashed-dotted line shows the limits without GDE but 1\,\% systematic uncertainty. 
We compare our limits 
to previous results for the same observation time 
but not including systematic uncertainties and GDE~\cite{Silverwood:2014yza,Lefranc:2015pza,Carr:2015hta}.
The annihilation cross section resulting in the right relic density 
is shown as a gray dashed line using the result from 
Ref.~\cite{Steigman:2012nb} (and an extrapolation thereof to 100\,TeV).
The NLO approximation of \textsc{PPPC4DMID} breaks down for 
$m_\chi > 10\,$TeV, indicated by the gray shaded region.}
\end{figure}

Before comparing the potential limits on the annihilation cross section 
in the EFT and simplified model frameworks to DD and LHC results, 
we compare our results for an annihilation of Majorana DM [$x = 1$ in Eq. \eqref{eq:annflux}] into 
$b\bar{b}$ quarks with the results of previous CTA sensitivity estimates \cite{Silverwood:2014yza,Lefranc:2015pza,Carr:2015hta}
(Fig.~\ref{fig:bb}). 
We assume a 100\,hour observation in the case of the morphological 
analysis, and 100\,hours for each RoI in the case of the true on/off
observations. 
The limits for our assumed Einasto profile
are an order of magnitude weaker than those assuming the 
NFW profile despite the larger observation time, due to the lower J~factor (see Fig.~\ref{fig:J}). 
For simplicity, we only show curves from other works 
that neglect systematic uncertainties and the GDE. 
This is also the main reason why, for the NFW profile we consider, our projected limits are 
worse by more than 1 order of magnitude.
If we also neglect both the effects of systematic uncertainties and the GDE, our limits improve by a factor of $\sim 12$ 
(blue dashed line in Fig.~\ref{fig:bb}).
In addition, if we use the DM density profile of Ref.~\cite{Pieri:2009je}, 
the limits improve by an overall factor of $\sim  20$
(light-blue dashed line in Fig.~\ref{fig:bb})
compared to our fiducial setup and are, in this case, comparable to those of Ref.~\cite{Carr:2015hta}.
We also show the limits if we neglect systematic uncertainties but include GDE (dotted blue line) and if we neglect GDE but include systematic uncertainties (dotted-dashed blue line). 
Inclusion of GDE has a large effect at high DM masses, since we have chosen a simple power-law extrapolation of the \fermiLAT GDE template which likely overestimates the GDE at high energies. 
Interestingly, the effect of systematic uncertainties dominates for $m_\chi \lesssim300$\,GeV in comparison to the GDE. 
The reason is the following
for low mass DM; only the first energy bins contribute to the likelihood due to the cutoff of the DM annihilation \gray spectrum. Furthermore, due to the smaller FOV of CTA at low energies, only the innermost spatial rings contribute. Yet, for these energy bins, the expected DM flux (for fixed $\langle\sigma v\rangle$) in each energy bin will be higher for a low mass DM particle compared to a high mass one since the DM flux is suppressed with $m_\chi^2$. In the likelihood fit, the relatively high expected DM flux for low masses can be compensated with the systematic uncertainty term (cf. Eq. \eqref{eq:lnl}). For high DM masses, the expected DM flux in each energy bin is small and hence the systematic uncertainty term has a smaller effect on the fit. However, more energy bins contribute to the the overall likelihood. Moreover, more spatial bins are included in the fit, further reducing the effect of systematic uncertainties.

We conclude that our analysis -- compared to previous analyses -- yields conservative 
results for the CTA sensitivity to the detection of DM 
due to the inclusion of systematic uncertainties,
 the GDE modeled without a high energy cutoff, 
and the lower J~factor.
We furthermore have not optimized the analysis in terms of the spatial or spectral binning which will be done in a forthcoming 
publication of the CTA consortium.

\subsection{Effective field theory}
\begin{figure}
\centering
\includegraphics[width = 0.9\linewidth]{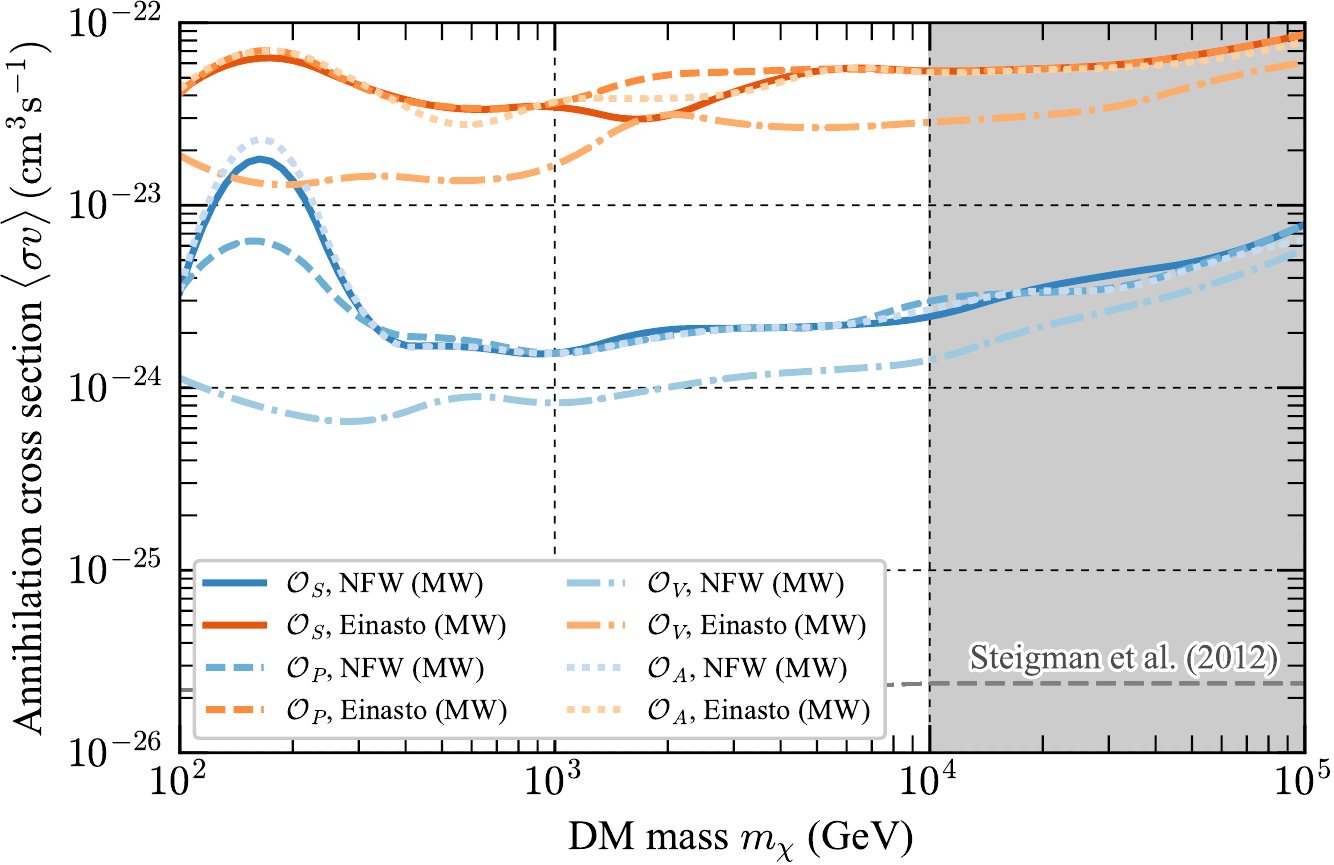}
\caption{\label{fig:EFTsv}Limits on the annihilation cross
section for the different EFT operators and DM density 
profiles. The NLO approximation of \textsc{PPPC4DMID} breaks down for 
$m_\chi > 10\,$TeV, indicated by the gray shaded region.}
\end{figure}

For each of the EFT operators listed in Sec.~\ref{sec:EFTtheory}, we derive 
upper limits on the cross section in the same way as 
with the pure annihilation into $b\bar{b}$ above. 
The results
are shown in Fig.~\ref{fig:EFTsv}. 
The assumed observation times are the same as before.
Remarkably, the limits are very similar for the $\mathcal{O}_{S,P,A}$ operators and are slightly better for the vector operator $\mathcal{O}_{V}$. This weak dependence on the exact operator 
demonstrates that limits from other operators not included in the present analysis should not yield very different results. 
As for the $b\bar{b}$ case, the limits degrade by an order of magnitude if the DM follows the Einasto profile (orange lines) instead of the NFW (blue lines). The weakening of the $\mathcal{O}_{S,P,A}$ limits at 180\,GeV appears when annihilation into top quarks becomes 
kinematically available and will be further discussed in the simplified model case below. 

\begin{figure*}[thb]
\centering
\includegraphics[width = 0.75\linewidth]{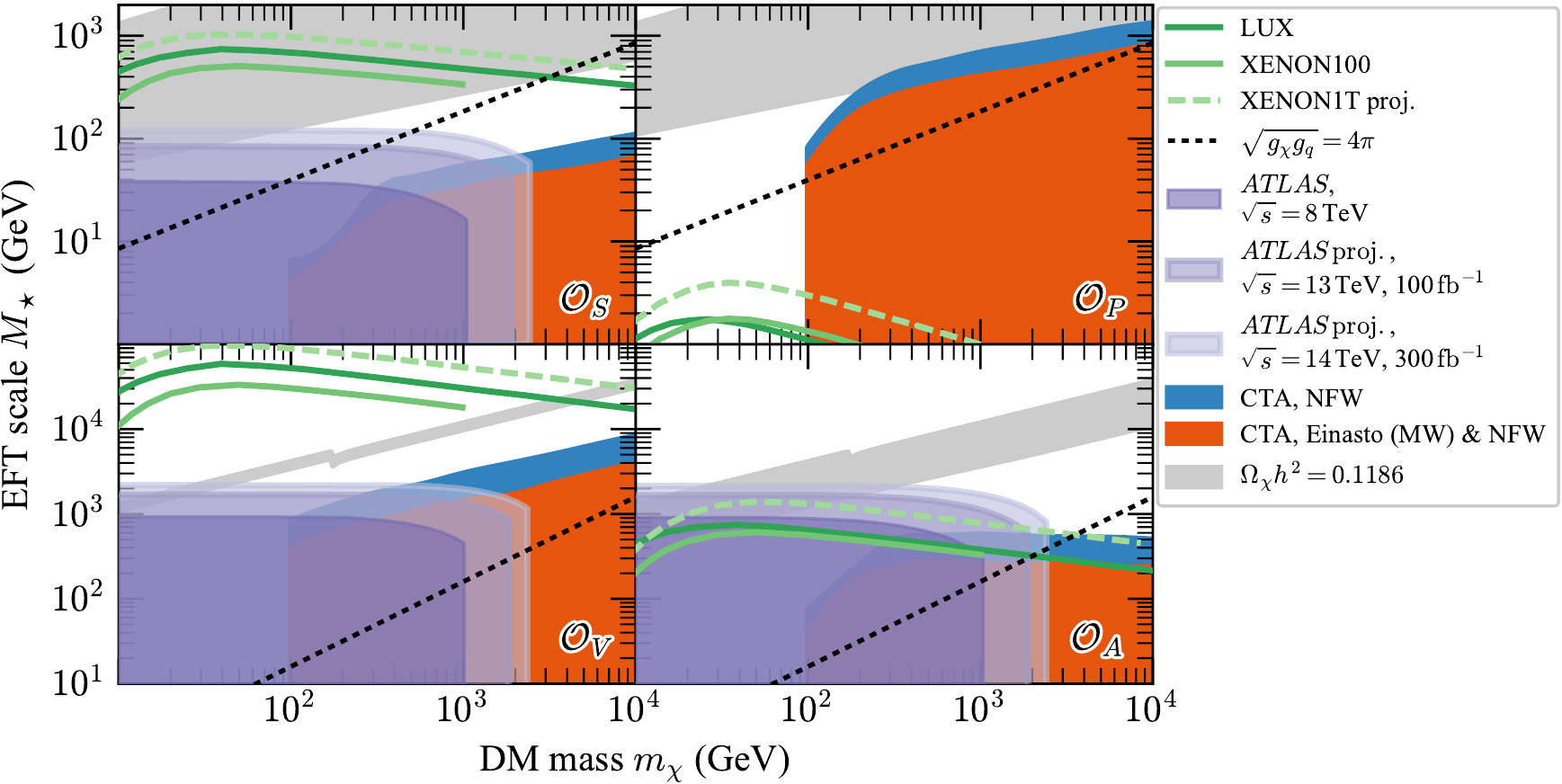}
\caption{\label{fig:EFTlimit} 
Lower limits on EFT scale $M_\star$ (red shaded regions) and comparison to LHC (purple shaded regions) and DD limits (green lines) for each tested EFT operator. Values of $M_\star$ and $m_\chi$ within the gray band result in the DM relic abundance as measured with \textit{Planck}
and the adopted range of values for $x_F$ and $g_*$.
The EFT approximation is valid above the black dotted lines. 
}
\end{figure*}

The limits on $\langle \sigma v \rangle$ can then be transformed into \emph{lower limits} on the 
EFT scale $M_\star$ using Eqs. \eqref{sigvOS}-\eqref{sigvOA}.
The constraints for our NFW and Einasto profiles 
are shown in Fig.~\ref{fig:EFTlimit} as dark-red and 
red shaded regions, respectively, together
with bounds from DD experiments (green lines) and 
the LHC (dark-purple shaded region). 
Due to the strong dependence of $\langle \sigma v \rangle$ on $M_\star$ ($\langle \sigma v \rangle \propto M_\star^{-6}$ for $\mathcal{O}_{S,P}$ and $\langle \sigma v \rangle \propto M_\star^{-4}$ for $\mathcal{O}_{V,A}$), the lower limits 
do not depend strongly on the assumed DM density profile in contrast to the limits on the annihilation cross section.

In general, the lower limits for CTA follow the 
expectations for the suppression of indirect detection of the operators listed in Table~\ref{tab:suppression}.
The strongest limits are found for the vector and 
pseudoscalar operators, $\mathcal{O}_V$ and $\mathcal{O}_P$, respectively.
Only for the pseudoscalar operator, CTA might be able to probe the cross section resulting in the correct thermal DM relic abundance and the corresponding values for $M_\star$.
These are given by the 
gray band in Fig.~\ref{fig:EFTlimit}. 
The band is derived from the standard equation~\cite{Gelmini:1990je},
\begin{equation}
\Omega_\chi h^2 = \frac{1.07\times10^9\,\mathrm{GeV}^{-1}}{M_\mathrm{Pl}}
\frac{x_F}{\sqrt{g_*}}\frac{1}{a + 3b/x_F},\label{eq:thermal}
\end{equation}
where $M_\mathrm{Pl} \approx 1.22\times10^{19}\,\mathrm{GeV} $ is the Planck mass, $h$ is Hubble parameter, $g_*$ is the number of relativistic degrees of freedom, and $x_F = m_\chi / T_F$ is 
the inverse freeze-out temperature scaled with WIMP mass. 
Following Ref.~\cite{Blumenthal:2014cwa} we take $20 < x_F < 30$~\cite{Jungman:1995df,Cao:2009uw} and $80 < g_* < 100$~\cite{Coleman:2003hs}. 
We emphasize that these choices are rather simplistic but are sufficient in the context of the EFT to estimate the relic density curves. 
A more accurate description is required once we get closer to more concrete model building realizations as done in the simplified dark matter model section that we will discuss below. In any case, the coefficients $a$ and $b$ stem from the expansion of the cross section, $\langle \sigma v \rangle \sim a + bv^2$, 
and can be read off directly from Eqs. \eqref{sigvOS}-\eqref{sigvOA}.
Setting $\Omega_\chi h^2 = 0.1$ as derived from \textit{Planck} measurements~\cite{Ade:2015xua}, the gray band follows from inserting the expressions for $a$ and $b$ into Eq.~\eqref{eq:thermal} and solving for $M_\star$. The band reflects the assumed range of values for $x_F$ and $g_*$.

\subsubsection{LHC comparison}
For comparison we show LHC constraints on EFTs from Ref.~\cite{Aad:2015zva} (shown as a dark-purple shaded region in Fig.~\ref{fig:EFTlimit}). EFT constraints at the LHC must be treated with caution, as the energy scale of the interaction may be large enough that the mediator is resolved, calling into question the validity of the EFT treatment. For recent reviews, see Refs.~\cite{Abdallah:2015ter,DeSimone:2016fbz,Kahlhoefer:2017dnp}. For the operators in question, the constraints are generally valid for effective coupling strengths of order unity or greater. Counterintuitively, the region of validity remains similar when moving from energy scales of 8 to 13 or 14\,TeV, since the baseline constraint on $M_\star$ is strengthened at the same time as larger mediator masses become accessible \cite{ATL-PHYS-PUB-2014-007}. We use the Collider Reach tool~\cite{collider-reach} to rescale the constraints from Ref.~\cite{Aad:2015zva} and provide an approximate estimate of prospective reach at center-of-mass energy 13\,TeV (14\,TeV) and luminosity 100\,fb$^{-1} ($300\,fb$^{-1}$) as light-purple shaded regions in Fig.~\ref{fig:EFTlimit}. These prospective limits should only be used as an indication, since the Collider Reach tool assumes that the details of the analysis are unchanged for the different center-of-mass energies and luminosities. 

Regardless of the assumed DM density profile, it is clear that CTA will play a complementary role in the search for dark matter. Moreover, it will be possible to probe higher DM masses compared to the LHC, even considering prospects at 14\,TeV and 300\,fb$^{-1}$.

Above $m_\chi \sim 1\,$TeV the lower limits 
from CTA are always more constraining than the limits 
from the LHC. Especially for the vector and pseudoscalar operators  CTA will be sensitive to DM annihilation 
signals out of reach of the LHC. The LHC should have a comparable sensitivity in the pseudoscalar case as in the scalar operator case \cite{Abercrombie:2015wmb}.

\subsubsection{Direct Detection}
DD limits are traditionally presented in terms of zero-momentum WIMP-nucleon cross sections. These are computed from WIMP-nucleon effective theories in which the WIMP interacts with nucleons via either a scalar operator $\bar{\chi}\chi\bar{N}N$ (``spin independent'') or an axial-vector operator $\bar{\chi}\gamma^\mu\gamma^5\chi\bar{N}\gamma_\mu\gamma_5 N$ (``spin dependent''), though recently some experiments have begun to adopt more general EFT schemes \cite{Schneck:2015eqa}. In order to compare these limits to those we compute for WIMP-quark effective operators, we need to relate the couplings of the WIMP-quark operators to those of WIMP-nucleon operators. We perform this translation using a common leading order 
prescription, recently reviewed in Ref.~\cite{DeSimone2016}. The four WIMP-nucleon operators that arise from the WIMP-quark operators we consider are then
\begin{eqnarray}
  c^{N}_S \mathcal{O}^{N}_S &=& c^{N}_S \bar{\chi}\chi\bar{N}N \\
  c^{N}_P \mathcal{O}^{N}_P &=& c^{N}_P \bar{\chi}i\gamma_5\chi\bar{N}i\gamma_5 N \\
  c^{N}_V \mathcal{O}^{N}_V &=& c^{N}_V \bar{\chi}\gamma^\mu\chi\bar{N}\gamma_\mu N \\
  c^{N}_A \mathcal{O}^{N}_A &=& c^{N}_A \bar{\chi}\gamma^\mu\gamma_5\chi\bar{N}\gamma_\mu\gamma_5 N
\end{eqnarray} 
where the coefficients of these operators can be expressed in terms of the coefficients of our WIMP-quark EFT operators as
{
\allowdisplaybreaks
\begin{eqnarray}
  c^{N}_S &=& \frac{m_N}{M_\star^3}\left(\sum_{q=u,d,s}f_q^{(N)} + \frac{2}{9}f_G^{(N)}\right) \\
  c^{N}_P &=& \frac{m_N}{M_\star^3}\sum_{q=u,d,s} \left(1 - \frac{6\bar{m}}{m_q}\right)\Delta_q^{(N)} \\
  c^{N}_V &=& \frac{3}{M_\star^2} \\
  c^{N}_A &=& \frac{1}{M_\star^2}\sum_q \Delta_q^{(N)}
\end{eqnarray}
}
where $\bar{m}=(1/m_u+1/m_d+1/m_s)^{-1}$, and  $f_q^{(N)}$, $f_G^{(N)}$, and $\Delta_q^{(N)}$ are experimentally measured quark-nucleon form factors, whose values we take to be the defaults from \textsc{DarkSUSY}~\cite{Gondolo:2004sc}. There is some uncertainty in these values, however the precise choice does not strongly affect our results. 

Next we need to predict a zero-momentum WIMP-nucleon cross section as is typically used by experiments. Following again Ref.~\cite{DeSimone2016} we can predict a SI cross section from $\mathcal{O}^{N}_S$ and $\mathcal{O}^{N}_V$, and predict a SD cross section from $\mathcal{O}^{N}_A$, according to
\begin{eqnarray}
\sigma_{SI} &=&\frac{\mu^2_{\chi{N}}}{\pi}(c_i^{N})^2 \quad \text{for }i=S,V, \\
\sigma_{SD} &=& \frac{3\mu^2_{\chi{N}}}{\pi}(c_A^{N})^2,
\end{eqnarray}
where $\mu_{\chi{N}}$ is the WIMP-nucleon reduced mass.
These predictions can then be compared directly to limits produced by DD experiments, which we translate back into a limit on $M_\star$. Strictly speaking the experimental limits are produced for some fixed DM halo model, generally an isothermal halo with some escape velocity, which complicates the comparison, but DD limits are generally not highly sensitive to the chosen halo model.
Yet, it is at least simple to rescale limits to suit a different local DM density. We do this where needed in order to match the value we use elsewhere in this analysis ($\rho_{local}=0.42$ $\mathrm{GeV}/\mathrm{cm^2}$).

The pseudoscalar case $\mathcal{O}^{N}_P$ is more difficult, because the nonrelativistic reduction of this operator does not coincide with either of the standard SI or SD operators used by experiments. The experimental constraints can therefore not be translated directly; one needs to reinterpret them by generating full predictions for the spectrum of recoil events that should be observed. We do not undertake this exercise; however the authors of \cite{DelNobile:2013sia} have done so, and have produced limits directly on the coupling $c^{N}_P$ using the same choice of WIMP-quark coupling structure as us, so we can directly use their translations of the experimental limits. These limits (originating from Refs.~\cite{Aprile:2012nq,Akerib:2013tjd}) are not quite as up to date as the ones we compute for the other EFT operators (based on Refs.~\cite{Aprile:2016swn,Akerib:2016vxi} for SI and Refs.~\cite{Aprile:2016swn,Akerib:2016lao} for SD), however they give a good idea of the current reach of the experiments. In particular $\mathcal{O}^{N}_P$ is momentum suppressed, and we see this in the weaker limits from DD experiments.

The resulting limits on $M_\star$ for XENON\,100 and LUX are shown as green lines in Fig.~\ref{fig:EFTlimit}.
In the case of a continued nondetection with these experiments,
these limits are likely to improve in the near future as the current generation of DD experiments such as XENON\,1T~\cite{Aprile:2015uzo} are currently taking data.
We show projections for XENON\,1T with a 2 ton-year exposure as a green dashed line
in Fig.~\ref{fig:EFTlimit}. 
These projections are derived by simply taking the fraction in between the input limits used in Ref.~\cite{DelNobile:2013sia} and the sensitivity of XENON\,1T \cite{Aprile:2015uzo} and multiplying the results of Ref.~\cite{DelNobile:2013sia} with the same fraction, working in the high WIMP mass limit.
This procedure of course ignores various details related to the spectral differences and assumptions about the future signal region, but should give a reasonable estimate of the reach for high WIMP masses.

In general, for the unsuppressed scalar and vector operators \cite{Busoni:2014sya} the measurements of these dedicated DM experiments result in more constraining limits 
than what can be expected from CTA. On the other hand, for operators $\mathcal{O}_P$ and $\mathcal{O}_A$ where we expect a suppression of the DD limits (cf. Table~\ref{tab:suppression}), CTA observations
will be able to yield complementary results.
For $\mathcal{O}_A$, this is the case for masses $m_\chi \gtrsim 1\,$TeV, whereas for the pseudoscalar case 
CTA limits will dominate over the entire tested DM mass range.

\subsubsection{EFT validity} 
The EFT approximation assumes that the energy scale of the underlying model cannot be resolved by the interactions under study. That is, for tree-level $s$-channel interactions, $M_{\rm med} \gg \sqrt{s}$. For the case of indirect detection where the annihilating DM is nonrelativistic $\left[s = 4m_\chi^2 + o(v^2)\right]$, this amounts to a requirement that $M_{\rm med} \gg 2 m_\chi$, assuming an $s$-channel underlying model.

For the $\mathcal{O}_V$ and $\mathcal{O}_A$ operators the connection between the mediator mass and the EFT scale is straightforward, 
$M_\star^{-2} \equiv g_q g_\chi /M_\mathrm{med}^2$.
Therefore, the EFT approach for these operators is valid as long as
\begin{equation}
M_\star > \frac{2m_\chi}{\sqrt{g_g g_\chi}}.
\end{equation}
For operators $\mathcal{O}_S$ and $\mathcal{O}_P$ the connection is more complicated, $m_q / M_\star^3 \equiv (m_q / \vev) g_qg_\chi / M_\mathrm{med}^2$, 
so that the validity condition reads \cite{Aad:2015zva}
\begin{equation}
M_\star > \left(\frac{1}{\vev}\frac{2m_\chi}{\sqrt{g_q g_\chi}}\right)^{\frac{2}{3}}.
\end{equation}

When the limits on $M_\star$ are weak, the coupling strength has to be large in order for $M_{\rm med}$ to be sufficiently large that the EFT approximation holds. These EFTs are not UV-complete by construction, and for sufficiently large couplings $\gtrsim 4 \pi$ will violate perturbative unitarity. At this point the EFT approximation fundamentally breaks down and cannot be considered to give an accurate description of a physical model \cite{Abdallah:2015ter,DeSimone:2016fbz,Kahlhoefer:2017dnp}. This threshold is shown as a black dotted line 
in Fig.~\ref{fig:EFTlimit} (below this line EFT is not valid). 
This limitation is especially severe for the scalar operator where CTA can only 
limit the EFT scale in parts of the parameter space where the EFT approximation 
breaks down. It is evident that collider searches and especially DD experiments 
are better suited to search for this type of DM.
The situation is less severe for $\mathcal{O}_A$ and $\mathcal{O}_P$ where 
the limits are, e.g.,  valid up to DM masses of 2 and 20\,TeV for the NFW profile, respectively. 
In the vector operator case, the limits are valid over the entire region of the
parameter space.

\subsection{Simplified models}

\begin{figure*}[thb]
\centering
\includegraphics[width = 0.75\linewidth]{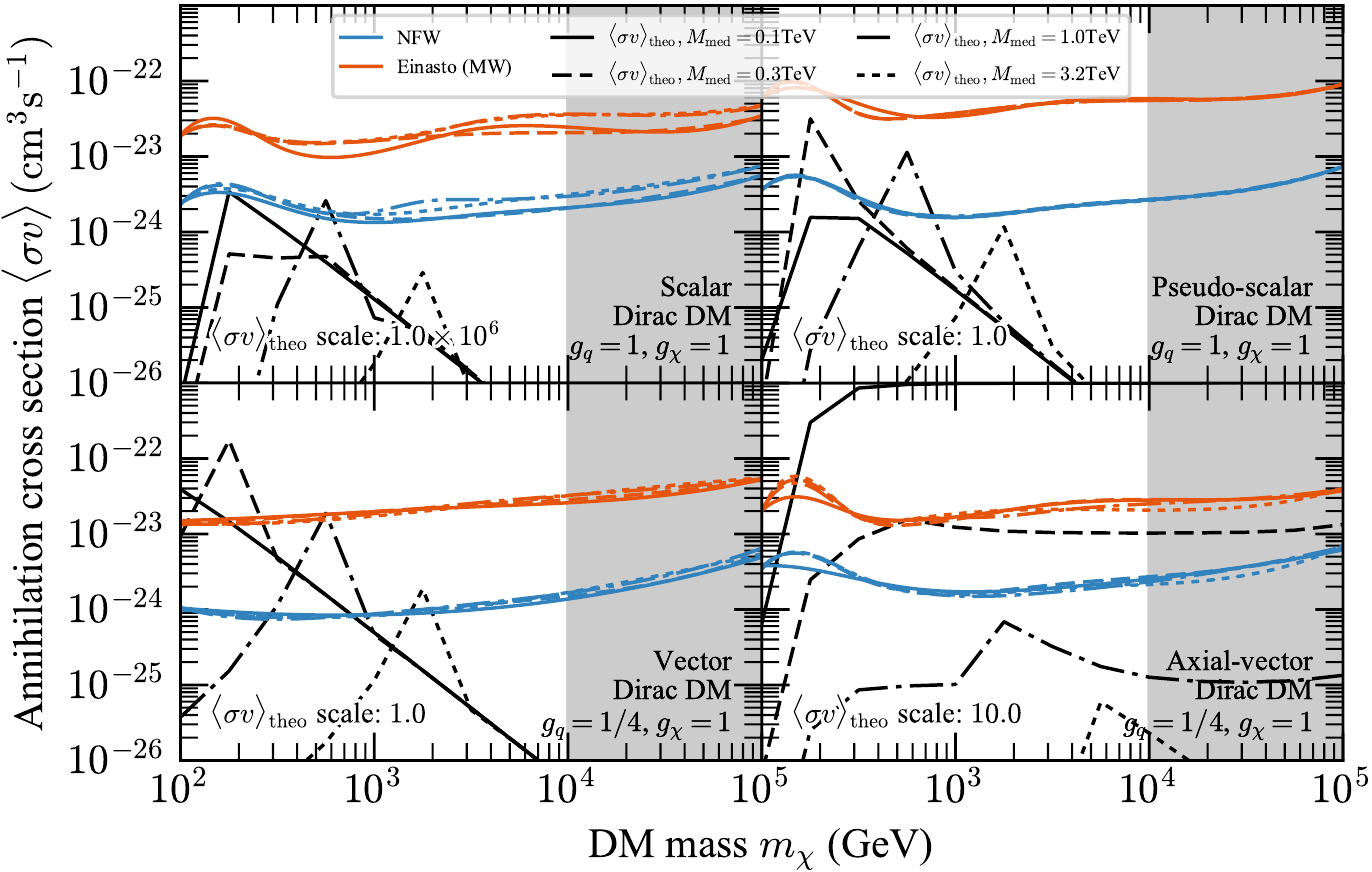}
\caption{\label{fig:SMcrossSection} Examples for excluded annihilation cross sections for the different simplified models. Limits are shown for both considered DM density profiles (blue and orange lines) and for different mediator masses $M_\mathrm{med} =$ 0.1, 0.3, 1, and 3.2\,TeV. Theoretical cross sections are also shown in black and are upscaled 
by a factor of $10^6$ and $10$ for the scalar and axial-vector DM, respectively. The NLO approximation of \textsc{PPPC4DMID} breaks down for 
$m_\chi > 10\,$TeV, indicated by the gray shaded region.}
\end{figure*}

We consider a logarithmic 13$\times$13 grid over the mediator and DM mass for 
each simplified model in the range of 100\,GeV and 100\,TeV. 
For each grid point we derive upper limits on the annihilation cross 
section in the same way as for the $b\bar{b}$ spectrum and the EFT operators. 
We show the limits on $\langle \sigma v \rangle$ for four mediator masses and
all considered values of $m_\chi$, the two DM density profiles, and operators in Fig.~\ref{fig:SMcrossSection}. 
In order to convert these limits into exclusion regions
in the $M_\mathrm{med}$-$m_\chi$ plane, we consider the theoretical 
values for the annihilation cross section, $\langle \sigma v \rangle_\mathrm{theo}$ for each pair of $M_\mathrm{med}$
and $m_\chi$, calculated through Eqs.~(\ref{sigvtotS})--(\ref{sigvtotA}).
The theoretical cross sections are shown as gray lines for each mediator mass 
in Fig.~\ref{fig:SMcrossSection}.
As anticipated from Table~\ref{tab:suppression}, these cross sections for the scalar and axial-vector DM case are suppressed and the values of
$\langle \sigma v \rangle_\mathrm{theo}$ are scaled upward for better visibility. 
For all simplified models apart from the vector DM one a bump in the 
limits is visible at $m_\chi = 0.18\,$TeV (same as in the EFT case).
As discussed in Secs.~\ref{sec:EFTtheory} and~\ref{sec:simpmods},
this feature arises when annihilation 
into $t\bar{t}$ quarks becomes kinematically accessible. The opening of this channel also leads to a jump in the hardness of the photon spectrum per annihilation, visible in Figs.~\ref{fig:specEFT} and~\ref{fig:specSM}.
Aside from these jumps, the annihilation rate falls off as $\sim m_\chi^{-2}$. For higher DM masses, the loss in sensitivity is remedied by a larger number of $\gamma$-ray energy bins that contribute to the likelihood in Eq. \eqref{eq:lnl}. 
This falloff with DM mass is not seen in the axial-vector case, an indication of the pathological behavior in the UV of this model, discussed further later in this section. 

\begin{figure*}[thb]
\centering
\includegraphics[width = 0.8\linewidth]{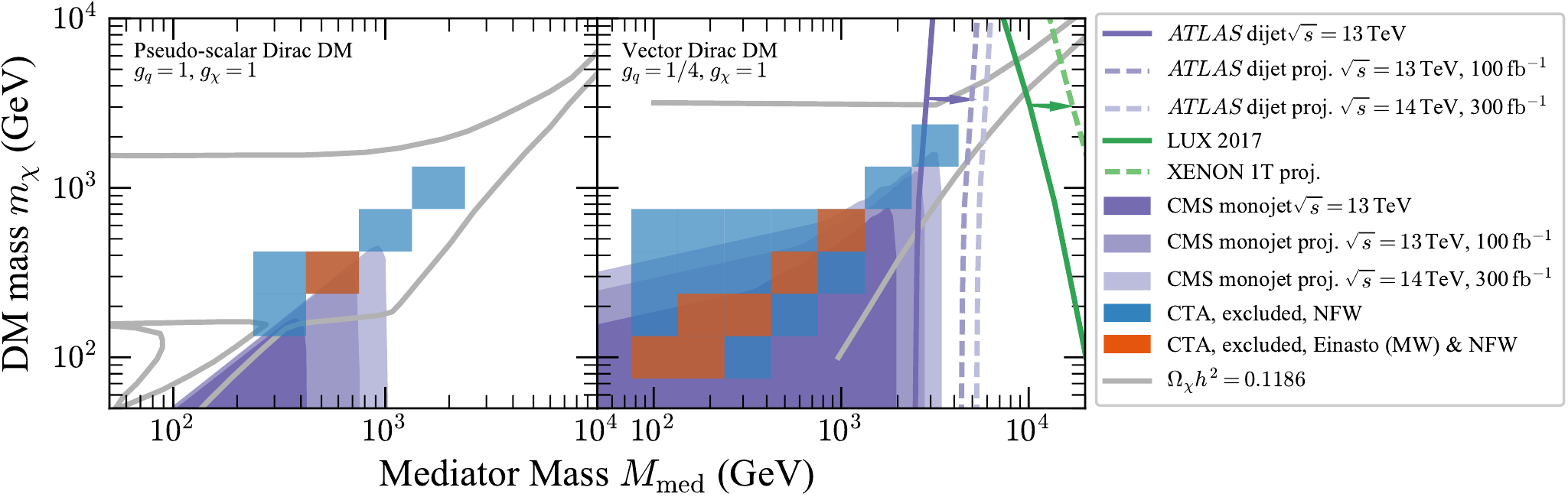}
\caption{\label{fig:SMlimit} Projected limits and excluded values of $m_\chi$ vs $M_\mathrm{med}$
for the different DM models. 
CTA observations will exclude combinations of $m_\chi$ and $M_\mathrm{med}$ indicated by the orange (NFW and Einasto DM profiles) and blue (NFW only) squares. 
Only the parameter space to the right of the green and purple solid lines is allowed from current DD and dijet limits, respectively (dashed lines show projections of future searches). Purple regions show LHC monojet limits and projections.
The gray lines show the parameter values that yield the correct relic abundance. 
No points are excluded in the scalar Dirac DM case for 
$g_q = g_\chi = 1$ and none of the excluded parameters of the axial-vector model obey perturbative unitarity, thus these cases are not shown. }
\end{figure*}

For the points in the parameter space where $\langle \sigma v \rangle_\mathrm{theo}$ is larger than the limits on the cross section, these particular combinations 
of $M_\mathrm{med}$ and $m_\chi$ are ruled out (e.g. $m_\chi = 0.2\,$TeV and 
$M_\mathrm{med} = 0.3\,$TeV for the pseudoscalar case and the NFW DM density profile). 
These excluded regions of the parameter space are shown in Fig.~\ref{fig:SMlimit} 
together with the combinations of $M_\mathrm{med}$ and $m_\chi$ that yield the correct relic abundance and 
limits from the LHC and DD. 
Only constraints for the pseudoscalar 
and vector DM models are presented.
For scalar DM, none of the tested mass points are ruled out, due to the strong suppression of the theoretical annihilation cross section.
In the EFT case this is reflected by the fact that none of the derived limits are in the EFT validity range.
In the axial-vector case, CTA observations only rule 
out models for which the mediator masses are small
but the DM mass is large. 
In this region, the model violates perturbative unitarity (see below).

\subsubsection{Relic Density}

For simplified models, the nonrelativistic limit of the relic density calculation employed for the EFT scenario is no longer accurate. The addition of a mediator particle causes the nonrelativistic approximation of the annihilation rate to break down around the resonant enhancement region ($M_{\rm med} \simeq 2m_\chi$) and at the threshold of mediator production becoming kinematically accessible ($M_{\rm med} \lesssim m_\chi$). 

The full relic density calculation entails solving the Boltzmann equation that determines the abundance of the DM particles at a given temperature, $Y(T)$, defined as the
number density divided by the entropy density as follows, \cite{Gelmini:1990je,Edsjo:1997bg}
\begin{equation}
 \frac{dY}{dT}= \sqrt{\frac{\pi  g_*(T) }{45}} M_\mathrm{Pl} \langle\sigma v\rangle(Y(T)^2-Y_\mathrm{eq}(T)^2)
    \label{eq:dydt}
\end{equation}
where $g_{*}(T)$ is the temperature-dependent effective number of degrees of freedom, 
 $Y_\mathrm{eq}(T)$ is the DM abundance in thermal equilibrium, and $\langle\sigma v\rangle$ is the relativistic thermally averaged DM annihilation cross section. The latter captures the specifics of each simplified model used here, including all possible annihilation channels. 
In the simplified models we study, we include only pair annihilations and no coannihilations, in which case the thermally averaged cross section is found to be~\cite{Gondolo:1990dk}
\begin{equation}
       \langle\sigma v\rangle=  \frac{\int_{4m^2}^\infty ds
K_1(\sqrt{s}/T) \sum\limits_{k,l}\sigma_{kl}(s-4m^2)\sqrt{s}}
                         {8 m^4 T \left(K_2(m/T)\right)^2 }\;,
\label{sigmav}
\end{equation}
where 
$\sigma_{kl}$ is the total cross section for annihilation of a pair of particles with masses $m$ into the final states $(k,l)$, and  $s$ is the
invariant center-of-mass energy of the incoming particles. 
For instance, in the nonrelativistic limit,
$\sqrt{s}$ is simply twice the DM mass. $K_1  (K_2)$ are the modified Bessel functions of order one (two). These modified Bessel functions arise as the result of the integrals involving the Boltzmann factors.

In order to compute the abundance of the DM particle today, $Y(T_0)$, we integrate Eq.~\eqref{eq:dydt} from $T=\infty$ to $T=T_0$, leading to,
\begin{eqnarray} 
\label{eq:omegah} \Omega_\chi h^2 &=& \frac{8 \pi}{3}
\frac{s(T_0)}{M_\mathrm{pl}^2 (100\,{\rm km\,s^{-1}\,Mpc^{-1}})^2} m_{\chi}Y(T_0) \nonumber\\
&\approx& 2.742 \times 10^8 \frac{m_{\chi}}{\mathrm{GeV}} Y(T_0)
\end{eqnarray} 
where 
$s(T_0)$ is the entropy density today determined by the \textit{Planck} Collaboration~\cite{Ade:2015xua}. The procedure described above is 
handled numerically within \textsc{micrOMEGAS}~\cite{Belanger:2006qa}.
The resulting regions in the $M_\mathrm{med}$-$m_\chi$ parameter space
that give the expected relic density are shown 
in Fig.~\ref{fig:SMlimit} as gray lines.

The annihilation cross section into SM fermions given in Eq.~\eqref{sigvqqP} is proportional to  $m_q^2/{\vev}^2$, whereas the annihilation into a pair of pseudoscalar fields does not depend on the quark masses. Therefore, in the former case annihilation into heavy quarks plays a crucial role, whereas in the latter the ratio $M_{\rm med}/m_{\chi}$ is the key quantity. With these features in mind one can understand the behavior of the curves for the relic density as shown in Fig.~\ref{fig:SMlimit}. One can see that when the DM mass becomes larger than the mediator mass, then the annihilation cross section in  Eq.~\eqref{sigvPP} simply depends on the ratio $M_{\rm med}/m_{\chi}$, explaining the behavior of the relic density curves. Using the same logic, when $m_{\chi} \gg M_{\rm med}$ the annihilation cross section in Eq.~\eqref{sigvPP}  becomes constant, explaining the horizontal lines for $M_{\rm med} < 1$\,TeV. The kinks exhibited by the relic density curves are a result of the top quark kinematic threshold. In other words,  when annihilation into the top quarks is kinematically accessible, a sharp boost in the cross section takes place as a direct consequence of the  $m_q^2/{\vev}^2$ enhancement.

In the vector mediator case, the DM annihilation cross section into SM fermions is very efficient, converse to the pseudoscalar case where there is a suppression  proportional to the vacuum expectation value, since the vector mediator interaction with SM fermions is dictated by gauge symmetries. When the DM mass is much larger than the mediator mass, the annihilation cross section into fermions simply scales with $g_q^2 g_\chi^2$, whereas the annihilation into the vector mediators goes with $g_{\chi}^4$. Hence, the annihilation cross section is constant since the couplings are fixed to be $g_q=0.25$, $g_\chi=1$, explaining the horizontal curves in Fig.~\ref{fig:SMlimit}. However, if $m_{\chi} \sim M_{\rm med}$, then annihilation into vector mediators becomes kinematically possible changing the overall shape of the annihilation cross section and relic density curves as can be seen in Fig.~\ref{fig:SMlimit}.  A key feature of the vector mediator scenario is the pronounced resonance that happens for $M_{\rm med} \sim 2 m_{\chi}$, which dominates the annihilation cross section then governed by the vector mediator decay width $\Gamma_{V, \rm tot}$. 

We emphasize that we have assumed that DM annihilates into quarks only to facilitate comparisons with collider searches. However, the inclusion of other final states such as leptons and gauge bosons, would yield different predictions for the annihilation rates and introduce additional free parameters.
This would also introduce a stronger dependence on a particular model.
The inclusion of extra interactions is beyond the scope of this work which is focused on complementarity among different DM searches.

 \subsubsection{Direct detection}

The DM-nucleon scattering in the nonrelativistic limit mediated by a pseudoscalar field leads to the spin-dependent momentum suppressed process. This momentum suppression arises when we match the quark-level matrix element with the nucleon-level matrix element in the nonrelativistic limit \cite{Freytsis:2010ne}. Assuming $M_{\rm med} \gg t$ (where $t$ is the usual Mandelstam variable) the Lagrangian for the pseudoscalar mediator leads to the following scattering cross section,

\begin{equation}
\sigma^{SD} = \frac{9}{4} \frac{f_N^2}{m_{\chi}^2 m_N^2}  \frac{g_{\chi}^2 \mu_{\chi N}^2}{M_{\rm med}^4}\, q^4
\label{SDpseudo}
\end{equation}
where $q$ is the momentum transfer, $N=n,p$, and,
\begin{eqnarray}
f_N&=& m_n \left[ \sum_{q=u,d,s} \frac{g_q}{{\vev}} \tilde{f_q} \right.\nonumber\\
&{}&-\left.\bar{m}\left(\sum_{q=u,d,s} \frac{\tilde{f_q}}{m_q}\right) \sum_{q=u,d,s,c,t,b} \frac{g_q}{{\vev}} \right]
\end{eqnarray}
with $\bar{m}=(1/m_u+1/m_d+1/m_s)^{-1}$ in agreement with Ref.~\cite{Freytsis:2010ne}, where $\tilde{f_u}=-0.44,\tilde{f_d}=0.84,\tilde{f_s}=-0.03$ \cite{Freytsis:2010ne,Cheng:2012qr,Dienes:2013xya}.
As in the EFT case, this expression can be directly compared 
to the limits reported by DD experiments. 
The momentum suppression in Eq.~\eqref{SDpseudo} causes the spin-dependent scattering cross section to lie orders of magnitude below current sensitivity. For couplings of order one and pseudoscalar masses above a few GeV, even the next generation of DD experiments will not furnish restrictive limits on the parameter space for the simplified model.

In contrast to the DM-nucleon scattering cross section,  the annihilation cross section for pseudoscalar interaction between DM and quarks is not momentum suppressed due to the presence of an $s$-wave term in the cross section.
This particular feature of the pseudoscalar interactions makes CTA 
well suited for these kinds of simplified DM models with the considered couplings,
possibly outperforming collider and DD methods, depending on the true DM density profile in the GC.
If the DM density follows the adopted NFW profile, it will also be possible to probe parameters that yield the correct DM  relic density. 

For the vector mediator case, things change dramatically, since the scattering is now spin independent and not velocity suppressed. The vector current is simply proportional to the number of valence quarks, and for this reason the calculation of WIMP-nucleon matrix element is not subject to large theoretical uncertainties \cite{Fitzpatrick:2012ix}. In the end the scattering cross section is found to be

\begin{equation}
\sigma^{SI}= \frac{ 9 \mu_{\chi N}^2 g_{\chi}^2 g_q^2}{\pi M_{\rm med}^4 }.
\end{equation}
The limits from Ref.~\cite{Akerib:2016vxi} and projections from Ref.~\cite{Aprile:2015uzo} are translated to limits in the $M_\mathrm{med}$-$m_\chi$ plane 
and shown as a green line in Fig.~\ref{fig:SMlimit}, excluding the region left of the line.
Since we are adopting $g_{\chi}=1$ and $g_{q}=0.25$, the scattering cross section is large, and heavy mediators are needed to circumvent the DD limits.

\subsubsection{LHC Constraints}

LHC constraints on simplified DM models stem from searches for large missing energy events produced alongside with a visible counterpart such as a jet, lepton, or photon. For this reason such searches are generally referred to as mono-X searches.  The properties of the model dictate which data set furnishes stronger limits. For the pseudoscalar mediator, monojet searches are the most restrictive. In  Fig.~\ref{fig:SMlimit} we show the monojet CMS constraints~\cite{Sirunyan:2017hci}. The LHC provides strong constraints at small masses but quickly loses sensitivity at higher energies (dark-purple shaded region in Fig.~\ref{fig:SMlimit}). We have used the Collider Reach tool~\cite{collider-reach} to estimate prospective reach at center-of-mass energy of 13\, and 14\,TeV and luminosities of 100 and 300\,fb$^{-1}$, respectively (light-purple shaded regions in Fig.~\ref{fig:SMlimit}). CTA limits will be complementary to these constraints  for mediator masses below 1\,TeV, while becoming the discovery probe for higher DM and mediator masses. 

As for the vector mediator cases, mono-X searches are no longer the most promising. In this case, searches for dijet resonances with large invariant mass are the most sensitive~\cite{Alves:2013tqa,Abdallah:2015ter,Chala:2015ama,Sirunyan:2016iap,Fairbairn:2016iuf}. By imposing hard cuts in the invariant mass of the dijet events, one can reduce the large background from quantum chromodynamics and effectively search for (axial-)vector mediators. Such probes are particularly sensitive to the coupling $g_q$ and mediator mass. For this reason the dijet limit~\cite{ATLAS:2016lvi} in Fig.~\ref{fig:SMlimit} (blue solid line and blue dashed line for the sensitivity estimates) is fairly independent of the DM mass.\footnote{The dijet limits are 
actually produced for the axial-vector case but should be the 
same for the vector mediator case.}
Notice that in our simplified models, the mediators neither couple to leptons, the Higgs, nor to gauge bosons. We do not expect significant changes in the collider bounds with the inclusion of these interactions. For instance, with the inclusion of interactions with leptons, both vector and axial-vector mediator cases would be subject to a stronger collider limit by less than a factor of 2 on the $Z^{\prime}$ mass \cite{Alves:2015pea}. This more restrictive bound would stem from resonance searches for dilepton final states at the LHC, which typically give rise to tighter constraints than the dijet one \cite{Alves:2015pea}.

\subsubsection{Unitarity/perturbativity}

The simplified model paradigm assumes that some unspecified UV completion assures the consistency of the model, providing a mechanism for mass generation and ensuring features such as gauge invariance. However, the axial-vector model includes processes which can violate gauge invariance and perturbative unitarity in certain regions of parameter space, such that any UV completion would fundamentally alter the phenomenology of the model in these regions \cite{Kahlhoefer:2015bea}. In order to ensure that the model does not violate perturbative unitarity, the following conditions must be met:

\begin{eqnarray}
m_{\chi, q} &\lesssim& \sqrt{\frac{\pi}{2}} \frac{\Mmed}{g_{\chi,q}},\label{unitarity1}\\
\sqrt{s} &<& \frac{\pi \Mmed^2}{g_\chi^2 m_\chi}.\label{unitarity2}
\end{eqnarray}
For ID, $\sqrt{s} \simeq 2m_\chi$, and Eq.~(\ref{unitarity2}) reduces to Eq.~(\ref{unitarity1}).
For the chosen values of $g_\chi$ and $g_q$, the criteria are not met by the combinations of $m_\chi$ and $M_\mathrm{med}$ that CTA observations could test.

\section{Conclusions}
\label{sec:concl}
In this article, we have compared the sensitivity of the future CTA to constrain annihilating DM with exclusions obtained from DD experiments and DM searches at the LHC.
This comparison has been achieved by utilizing the frameworks of 
 EFTs and simplified models. 
Our sensitivity projections are based on realistic IACT observation schemes of the Galactic center and test two different 
DM density profiles which are compatible with 
recent observations. 
They also incorporate contributions from Galactic 
diffuse emission and possible systematic uncertainties.

Within EFTs and simplified models, it is straightforward to compare the derived sensitivity
with limits and projections from DD experiments and collider searches for DM at the LHC. 
This is not the case for limits that 
are reported for a pure annihilation into one particular channel.
We have found that for DM mediators for which the annihilation 
is neither velocity nor helicity suppressed (pseudoscalar and 
vector mediators; cf. Table~\ref{tab:suppression}),
CTA will be able to probe regions of the parameter space 
out of reach for present and possibly even future collider searches
(Figs.~\ref{fig:EFTlimit} and~\ref{fig:SMlimit}). 
It will also be possible to probe parts of the parameter
space that results in the correct DM relic abundance. 
In the case of vector mediated DM, strong constraints 
already exist from DD experiments and LHC dijet analyses,
and CTA observations are unlikely to improve on already existing bounds, but it will still introduce a compelling and orthogonal probe to the model. The situation is however different if the DM mediator is a pseudoscalar. In this case, the scattering cross section 
is suppressed by a combination of the DM spin and spin of the nucleus. Indeed, the scattering cross section is spin dependent and momentum suppressed at the fourth power rendering DD bounds to be very suppressed. Each $\gamma_5$ matrix in the Lagrangian for the pseudo scalar model results in a momentum suppression, yielding a suppression proportional to $q^4$, where $q$ is the momentum transfer, in the WIMP-nucleon scattering. Moreover, because in this model the couplings with quarks have a Yukawa-like structure, suppressed by the fermion mass, dijet limits are not very competitive, and thus far there is no dijet limit from LHC for this simplified model.

For such DM models, CTA observations will be indispensable to probe higher values of DM (and mediator masses). 
In the EFT framework, the derived limits only depend weakly 
on the assumed DM density profile in the Milky Way due to strong dependence of the EFT scale $M_\star$ on the limits on the annihilation cross section. This is not the case for simplified models where the limits strongly degrade from the considered NFW to the Einasto density profile (cf. Fig.~\ref{fig:SMlimit}). 
To summarize, our results illustrate the need for different techniques (DD, ID, collider searches) to probe all possibilities 
of DM models.

We stress that all calculations presented here assume 100\,hours of observation time of the Galactic center (and additionally 200\,hours in the case of an Einasto density profile for independent background determination). Such an observational program should be completed within the first years of CTA operation. 
Therefore, the projected limits are bound to 
improve as CTA will continue to observe the GC beyond the first years of operation (the limits are expected to improve roughly with the square root of observation time). 
Furthermore, several analysis choices should be optimized 
in future analyses, such as the choice 
of the spectral and spatial binning, as well as the treatment of systematic uncertainties and GDE.
For the GDE, a simple power-law extrapolation 
of the template provided by the \emph{Fermi}-LAT Collaboration was used here that likely overestimates the GDE contribution at very high energies. 
A careful treatment of the GDE and optimization of the analysis parameters will be conducted in a forthcoming publication 
of the CTA consortium.

\begin{acknowledgments}
This paper has gone through internal review by the CTA Consortium. The authors would like to thank Torsten Bringmann, Gabrijela Zaharijas, Michael Daniel, Fabio Iocco, Javier Rico, and Fabio Zandanel for useful discussions and comments on the manuscript. 
J.C. is a Wallenberg Academy Fellow.
M.M. is a Feodor-Lynen Fellow and acknowledges support of the Alexander von Humboldt Foundation. 
M.A.S.C. is supported by the {\it Atracci\'on de Talento} Contract No. 2016-T1/TIC-1542 granted by the Comunidad de Madrid in Spain, 
and also partially supported by MINECO under grant FPA2015-65929-P (MINECO/FEDER, UE).  MASC also acknowledges the support of the Spanish MINECO's ``Centro de Excelencia Severo Ochoa'' Programme under Grant No. SEV-2012-0249, and the support of the Swedish Wenner-Gren Foundations to develop part of this research.  This work in part was supported by the Australian Research Council Centre of Excellence for Particle Physics at the Tera-scale (Grant No. CE110001004).
\end{acknowledgments}

\bibliographystyle{apsrev4-1}
\bibliography{cta_eft}

\end{document}